\documentclass[11pt]{article}  

\usepackage{amsmath,amsthm,latexsym,amssymb,amsfonts,epsfig}

\newtheorem{Definition}{Definition}[section]

\newcommand{\be}{\begin{equation}}
\newcommand{\ee}{\end{equation}}
\newcommand{\ba}{\begin{eqnarray}}
\newcommand{\ea}{\end{eqnarray}}

\title{{\sf (Quantum) reference frames, relational observables, 
gauge reduction and physical interpretation}}
\author{
{\sf T. Thiemann}$^1$\thanks{{\sf 
thomas.thiemann@gravity.fau.de}}\\
\\
{\sf $^1$ Inst. for Quantum Gravity, FAU Erlangen -- N\"urnberg,}\\
{\sf Staudtstr. 7, 91058 Erlangen, Germany}\\
}
\date{{\small\sf \today}}

\makeatletter
\@addtoreset{equation}{section}
\makeatother

\begin{document} 

\maketitle

{\sf

\begin{abstract}
Reference frames are simultaneously among the most elementary and most fundamental physical concepts as they
are the basis of the interface between theory and experiment. Without a thorough understanding 
of how to operationally define and translate a reference frame into mathematics, the 
physical interpretation of theory calculations in terms of observational data is not possible.

This becomes the more important for classical gauge systems where not all mathematical objects that 
enter into the a priori mathematical description are observationally defined. The situation is 
particularly challenging in General Relativity where spacetime coordinates, a primary tool to describe a reference 
frame, are subject to spacetime diffeomorphisms considered as gauge transformations which by definition
turns these into non-observables. 

As old as General Relativity itself is the idea of operationally defined (material) reference frames
which specify coordinates in terms of matter or geometry reference fields. 
This gives rise to the concept 
of relational observables and relational reference frames, 
e.g. the metric field relative to the electromagnetic field, which leads 
to a gauge reduction of the system. 

Upon quantisation, all fields become operator valued distributions. Now new conceptual and 
technical questions arise such as: Should one reduce before or after quantisation and 
how are the reference fields quantised respectively in either route? 
Is a reference frame itself subject to quantisation and how are different quantum
reference frames related? How does the gauge reduction fit into this, i.e. how can  
it be that a certain reference field is considered a non-observable in one reference frame and 
an observable in another which upon quantisation even displays fluctuations? 
How precisely are gauge dependent fields interpreted in terms of the relational observables 
in a given reference frame? What is the relative dynamics, e.g. how exactly are physical 
Hamiltonians of two relational reference frames related?  

In the present conceptual work we will address these and related questions in 
a non-perturbative field theory context of sufficient generality to cover General Relativity coupled 
to standard matter and aim for a hopefully consistent picture. A central role is played by 
the concept of the relational reference 
frame transformation (RRFT) for which a general formula is derived and its properties are 
explored.
\end{abstract}

\section{Introduction}
\label{s1}

Reference frames lie at the very heart of theoretical and experimental physics.
The mathematical framework of the theory has to be formulated using those space and 
time coordinates in which the laboratory produces the data in order to be able to 
directly and unambiguously relate computational numbers to measurement data. Furthermore, measurements 
themselves maybe required in order to determine in which reference frame the laboratory 
prevails. 

This issue becomes 
more and more important and ever more complicated as one increases the complexity of the theory. 
In classical non-relativistic mechanics 
the Lagrangian must be covariant with respect to Galilei transformations 
between inertial frames in order that physical equations of motion take their standard form. 
Here an inertial frame is such that particles move on straight 
lines when no forces act on it, an intrinsically operational definition which can be checked 
by performing experiments. In special relativistic mechanics, the Galilei group is replaced by 
the Poincar\'e group and in addition one faces the issue of gauge invariance for the 
first time: Not all spacetime coordinates that describe the trajectory of a free particle
are observable but only the relation of say values of the spatial coordinates to the value of the 
temporal coordinate. The underlying gauge symmetry is reparametrisation covariance of the
relativistic free particle Lagrangian or equivalently the corresponding Hamiltonian 
is constrained to vanish by the mass shell condition on the momenta conjugate to the 
particle. In passing to special relativistic field theory like the standard model 
of elementary particle physics on Minkowski space, the Poincar\'e group is a Killing symmetry of the 
Minkowski metric and inertial frames are again important in order that Euler-Lagrange 
equations take their standard form. In such theories, spacetime coordinates are still 
assigned the status of labels of the field, i.e. the spacetime position at which fields 
take certain values and these values can be assigned independently to all fields and their 
velocities on initial data surfaces. In passing to general relativistic field theories 
the initial data are no longer free. There are initial value constraints which have their 
origin in the diffeomorphism covariance of the underlying Lagrangian. This means that the spacetime 
coordinates that one uses to define the mathematical fields are no longer directly observable, very similar 
to the free relativistic particle. In particular,
changes with respect to a mathematical temporal coordinate are not directly observable, hence 
misinterpreting these as dynamics of observables would lead to wrong conclusion that 
observables should be frozen in General Relativity. What is again observable are relations of field values,
say the value of the metric field at that mathematical coordinate at which four matter 
fields or four curvature scalars take certain values. One calls such fields to define 
coordinates material or geometrical reference fields or ''clocks'' \cite{0,1}. 

As nature is general relativistically covariant as far as present experiments can tell, the 
latter most complicated situation is in fact the one that one must deal with. While the above idea of using 
material or geometrical clocks is as as old as General Relativity itself and intuitively
appealing, conceptually and technically it is quite involved and gives rise to several 
subtleties and puzzles which motivated the present work:
\begin{itemize}
\item[1.] The map between field values 
and spacetime coordinates may not be a diffeomorphism by itself, thus the relational observable
becomes ill defined. For instance using geometrical clocks in terms of curvature scalars, the spacetime coordinates in 
regions where the Riemann tensor vanishes cannot be unambiguously identified.
\item[2.]
Even if that complication can be avoided, e.g. by restricting the mathematical field
configurations of the reference fields to be valued in the diffeomorphism group, 
one is free to choose the reference fields as the theory itself does not prescribe how to 
choose them. How does the mathematical theory depend on that choice and what happens 
under change of reference fields? As the reference fields are used to identify spacetime 
coordinates, we take the notion of reference fields as synonymous with the notion of 
a reference frame. Thus we ask what happens under change of reference frames. 
\item[3.] 
By definition, the reference fields acquire the status of gauge degrees 
of freedom because by performing unobservable, mathematical coordinate transformations (passive 
diffeomorphisms), that are considered gauge transformations
in General Relativity, we may change their values quite arbitrarily. Gauge fields, i.e. the 
reference fields, are 
not observable. What is observable are the corresponding relationally defined fields, the 
relational observables.
This is crudely analogous to the Higgs mechanism: The mathematical formulation starts with 
a complex Higgs iso-dublett or a real quartet but by isospin gauge invariance three of these 
real fields are not observable. Using these as reference fields leaves one real Higgs Higgs scalar and three 
(massive) vector bosons as relational observables. One expects a similar 
mechanism to work for the diffeomorphism gauge group replacing the isospin gauge group.
Now in General Relativity, in an experiment one could in principle 
almost non disturbingly measure all ten components of the metric tensor in a given spacetime region by studying a geodesic 
congruence of almost massless, very slow (test) particles. How can that be if by above mechanism 
only two polarisations of gravitational 
fields are observable? 
\item[4.] Now consider change of reference fields.
Then different mathematical objects are assigned the label ''observable´´ and not "not observable".
How does this change of reference frame manifest itself in the reinterpretation of 
observational data in General Relativity? Does it mean that different components of the 
metric tensor become observable or not observable by changing the role of reference fields?
\item[5.] Surely, in some very concrete sense, the physics of the system must be 
independent of the choice of reference fields because the theory did not not know about this 
choice to be begin with. Is a change of reference frame therefore simply a symmetry transformation 
under which the equations of motion for the observables, are covariant? For instance, in the 
Hamiltonian formulation, are these simply canonical transformations that preserve the Hamiltonian 
which drives the dynamics of the respective observables?
\item[6.] In more technical terms, a choice of reference frame, as it appears in the 
present context, is a choice of reference 
fields together with a gauge fixing condition on them,
namely one imposes that the reference fields take certain coordinate value. Now using these and the 
initial value constraints, within the Hamiltonian formulation, one can in principle write a 
concrete formula for a relational observable \cite{3c,3d} and a concrete formula for the 
physical Hamiltonian, itself a relational observable, that drives their dynamics. It turns out 
that there is a Poisson isomorphism (canonical transformation) between this manifestly gauge 
invariant formulation and the so-called reduced phase formulation in which one identifies 
the relational observables with the true degrees of freedom which are those fields that one did
not solve the constraints and gauge fixing conditions for. Under this identification 
the physical Hamiltonian becomes the reduced Hamiltonian which only depends on the true degrees of 
freedom. Therefore a choice of reference frame is equivalent to
a choice of reduced phase space formulation. Again we can ask how a change of reference 
frame manifests itself at the level of the reduced phase spaces and reduced Hamiltonians.  
\item[7.] Suppose a gauge fixing condition was chosen, say that certain components of 
the metric tensor take certain values. What if the experiment contradicts that gauge 
fixing condition? How can one then match observational data to theory computations?
\item[8.] These issues become particularly important when we pass to the quantum theory. 
Here we have the choice between various quantisation schemes. Within the Hamiltonian 
setting we can use quantisation before and after reduction. Quantisation before reduction means that 
one imposes quantised versions of the constraints as generalised zero eigenvalue equations on 
physical states, defines a physical inner product among them and defines quantised versions 
of relational observables as self-adjoint operators on the resulting physical Hilbert space.
After reduction means that one quantises only the reduced phase space to begin with and 
interprets the corresponding Hilbert space as physical Hilbert space. It is by no means 
granted that these two procedures commute although one expects that their semiclassical limits 
agree.
\item[9.] Consider the reduced phase space route and reduction before quantisation for concreteness. 
Consider again a change of reference frame. How do the corresponding quantum theories change? Is it just a unitary 
transformation? As in different reference 
frames different classical fields acquire the status of non-trivial observables while the reference fields become 
just real numbers times the unit function, their quantisations yield non-trivial and trivial operators (proportional to the unit operator) respectively.
This means that changing reference fields also means that different fields are being (non-)trivially quantised.
This can therefore be considered also as version of a change of quantum reference frame. 
One sees that the unitary transformation, if it exists, cannot be as trivial merely exchanging 
canonical pairs of observable and non-observable fields because that would mean that in one frame 
the corresponding field is fluctuating while in the other it is not. This would have observational 
consequences and would mean that a reference frame is selected. What is the resolution of this
puzzle?
\end{itemize}
In this paper we attempt to answer those and related questions or at least suggest a strategy 
for finding an answer within a concrete mathematical framework, namely the 
reduction before quantisation approach. This explicit framework has been 
developed and applied by various authors in multiple contexts. To the best of our knowledge
the historically first appearance is in \cite{3a} in the context of gauge unfixing. In the 
context of relational quantum mechanics it was introduced in \cite{3b}. The generalisation 
to field theories was developed in \cite{3c} and the correspondence between reduced phase
space and relational observables of field theories was established in \cite{3d} which is closest to the 
notation that we will use here. Examples for applications to full classical or quantum 
General relativity using various kinds 
of geometrical or material reference systems in classical and quantum gravity are \cite{3e}. 
Also there is by now a rich literature on applications in symmetry reduced situations such as 
quantum cosmology and quantum black holes, see the relevant chapters in \cite{3f}. For the he most concrete application 
of this framework to (quantum) reference frames in the context of quantum gravity 
see \cite{3g} and references therein. An excellent textbook on quantum gauge reduction is \cite{5}. 

Of course we are touching only
on a tiny aspect of the vast field of (quantum) reference frames. This is a research topic 
of growing interest across several communities. One of the earliest 
appearances is \cite{4a}. The subject started to gain wider attention from 
\cite{4b}. The information theoretic aspect is emphasised in \cite{4c}. Applications 
to local quantum field theory on background spacetimes with symmetries are found in \cite{4d}.
In this work we will consider particular types of questions that arise in the quantum gravity role of 
quantum reference frames, see \cite{3g} and references therein for important contributions to the route. 
We will focus on concepts rather than deeper mathematical aspects. 
However, we will mention out subtleties with particular constructions 
and point to more detailed analysis along the way. For instance, most considerations will 
be local and more global analysis is required to complete the picture. \\
\\
This contribution is organised as follows:\\
\\

In section \ref{s2} we will briefly review the mathematically equivalent relational and reduced phase space formulations of 
constrained systems and the corresponding physical and reduced Hamiltonians respectively, 
to make this article self-contained and to fix the terminology. Most details can be found in \cite{3c,3d}.
We connect to the research field of reference frames as considering a choice of reference frame as synonymous with a choice of 
reference fields and gauge conditions on them, which in turn define a choice of relational observables or equivalently 
true degrees of freedom.   

In section \ref{s3} we define the central tool of the present article, namely the \{it relational reference frame transformation} within our concrete 
framework. We will not consider the most general situation but it will be general 
enough to put the finger on above and related questions. A similar concept is considered 
also in \cite{3g} but there with restrictions on the way the reference field is coupled 
to the constraints, technically it covers only the case of a deparametrised theory with finitely 
many degrees of freedom of which
General Relativity coupled to standard matter is not an example. Also changes of reference frames 
are considered in \cite{3g} but only for the very special case that the reference fields that are 
exchanged have no self-interaction and no interactions with any other field.
Our definition is free from both assumptions but  
it carries subtleties due to its local character.
We will show that it is a two parameter family of canonical transformations between the relational 
observables defined by the two reference frames.

In section \ref{s5} we offer a geometrical or graphical picture of the relational reference frame transformation 
for a system with a four dimensional kinematical phase space and single constraint to fix the intuition.

In section \ref{s3a} we will 
show in concrete examples that the physical Hamiltonians corresponding 
to two different choices of reference frames generically drastically differ from each other 
in the sense that when pulling back the physical Hamiltonian of the second reference frame
by the relational reference frame transformation one does not obtain the physical Hamiltonian 
of the first frame. Moreover, the difference between the resulting expression and the 
Hamiltonian of the first frame is generically a rather non-linear, even non-polynomial 
function of the relational observables of the first reference frame. 
This may be counter intuitive because one might have 
expected that the physical Hamiltonian, by itself a gauge invariant observable, should 
be frame independent, given that a change of reference frame in our definition means changing 
the reference field and/or the gauge condition, i.e. is very closely related to a gauge transformation
and gauge transformations do not change observables. Conversely, it is often criticised 
that the dependence of the Hamiltonian on the choice of gauge reduction indicates a loss 
of gauge invariance. We show that this is not the case and explain why this non-trivial transformation 
law is expected for various reasons. To avoid confusion, note that the formalism directly produces the 
general formula for the physical 
Hamiltonian in terms of the relational observables of a given relational reference frame, 
it does not need to be derived from the relational reference frame transformation.
 
In section \ref{s4} we resolve the fluctuation paradox. This 
is the observation that the relational observables corresponding to the 
reference fields that define a reference frame are quantised as 
multiples of the unit operator while with respect to another reference frame they are non-trivial 
operators. We then explain how relational quantum clocks fit into this picture that have non-trivial and observable
fluctuations in any reference frame. We also comment on the rather different role that kinematical clocks
play in the quantisation before reduction route. 

In section \ref{s6} we explain how a given function on the kinematical phase space is expressed 
in terms of the relational observables with respect to a choice of reference frame. We also 
prescribe what to do when computational and laboratory reference frame are misaligned and what
role is played by the relational reference frame transformation in that procedure.

In section \ref{s7} we sketch some of the technical challenges that have to be 
faced when trying to quantise the relative reference frame exchange map which up to this 
point  was 
defined only in the classical theory. This also involves to globalise the constructions made 
up to this point for which we sketch some ideas. We also describe how the quantum relational reference transformation
fits in with the bulk on the literature on quantum reference frames such as \cite{4b}.

In section \ref{s8} we summarise the findings of the main text including (partial) answers to above 
questions in non-technical terms and 
at the same time set up a glossary of the terminology used in this article. Readers not interested 
in the technical details of the article may jump immediately to that section.

In appendix \ref{sa} we embed field theories in Minkowski space into the framework 
of generally covariant gauge systems and apply the relational reference frame transformation 
corresponding to changes in just gauge fixing conditions on the same reference fields which,
in the simplest case, 
can be interpreted as a change between inertial frames. We recover the well known fact 
that the Hamiltonian depends on the inertial frame. This construction has appeared 
in parts already in \cite{4e}. What is different is that we work in any dimension and 
the quantisation after reduction route and what is new is that we systematically derive the 
quantum relational reference frame transformation in this case from the general formalism 
developed in the main text.

\section{Relational and reduced phase space formulation of constrained systems}
\label{s2}
   
We consider a Hamiltonian system with canonically conjugate phase space coordinates $(K^A,M_A), (X^I,Y_I)$.
In field theory the indices $A,I$ run through a countably infinite index set, in mechanics it is a finite set.
In field theory such a discrete labelling of degrees of freedom rather than by continuous points on a 
hypersurface can be achieved by writing the fields in terms of their mode components 
with respect to a real orthonormal basis of the Hilbert space of square integrable functions the hypersurface.
See the appendix for a concrete example.
It is important to stress that in this article we therefore are not limited to finitely many degrees of 
freedom but cover the field theory case.
 The phase space $\Gamma$ coordinatised by $K,M,X,Y$ is called the kinematical phase space. It is constrained 
by initial value constraints $Z_I(X,Y)=Y_I,\;C_I(K,M)$ which are first class in Dirac's terminology \cite{6}, that is,
$\{C_I,C_J\}=\kappa_{IJ}^K\; C_K$ for certain functions $\kappa$ of $K,M$ called structure functions.
Note that trivially $\{Z_I,Z_J\}=\{Z_I,C_J\}=0$. 

The Hamiltonian of the theory is totally constrained, i.e. is a linear combination of constraints
\be \label{2.1}
H=V^I\;Z_I+X^I\; C_I
\ee
This is the form of the Hamiltonian of the standard model of elementary particle physics coupled to 
General Relativity. The meaning of the various fields is as follows: The Lagrangian $L$ of the system 
is of the form $L(K,U;X,V)$ where $U,V$ are the velocities of $K,X$ respectively. To pass to the 
Hamiltonian formulation, one performs the Legendre transform by introducing the conjugate momenta 
$Y_I=\frac{\partial L}{\partial V^I},\;  P_A=\frac{\partial L}{\partial U^A}$, solving those equations for 
$U,V$ in terms of $K,M,X,Y$ and substituting this into $V^I\;Y_I+ U^A\;P_A-L$ resulting in $H(Q,P,X,Y)$.
This would be the end of it if solving for the velocities was possible. However, in gauge systems 
the Legendre transform is singular and one cannot solve for all velocities. In the simplest case such a 
singular Lagrangian has the form $L(K,U;X)$ which does not depend on $V$ at all. Then the Legendre transform 
is incomplete, $V$ remains undetermined and the Hamiltonian assumes the form $H=V^I\; Z_I+H'(K,M;X)$
with $Z_I=Y_I$ and $Y_I=\frac{\partial L}{\partial V^I}=0$ as so called primary initial value constraints.
Dynamical stability of the primary constraints enforces the secondary constraints $\{H,Z_I\}=\frac{\partial L}{\partial X^I}=:C_I=0$.
In principle this calls now for tertiary constraints etc. but it turns out that for the theories of 
interest, this so-called Dirac algorithm stops at the secondary level. Moreover, $H'=X^I\;C_I$ 
which brings us into contact with (\ref{2.1}). The Hamiltonian equations of motion for 
$K,M,X,Y$ derived from $H$ together with $\frac{\partial H}{\partial V^I}=0$ are completely 
equivalent to the Euler Lagrange equations of $L$ with respect to $K,X$.
Concretely, the fields $X$ comprise temporal components of Yang-Mills 
potentials and temporal-temporal (lapse) and temporal-spatial (shift) components of the metric field \cite{7}.
The $C_I$ comprise Yang-Mills type Gauss constraints and spatial diffeomorphism and Hamiltonian constraints.

Now $H$ contains the completely unspecified functions $V^I$ which play the role of Lagrange 
multipliers enforcing $Z_I=0$. Let $t$ be the parameter of the foliation of spacetime into 
Cauchy surfaces (in mechanical systems the Cauchy surface is a point on the $t$ axis). Then 
the Hamiltonian equations of motion dictate  
$\frac{\partial X^I}{\partial t}=\{H,X^I\}=V^I$. Thus the velocity of $X^I$ is unpredictable, hence 
the field $X^I$ is unpredictable. Then $\{H,K^A\},\;\{H,M_A\}$ also contain $X^I$ and are thus also 
unpredictable. Thus one is forced to interpret the motions generated by $H$ as gauge transformations
and $t$ itself cannot have the usual status of the time parameter whose translations are 
generated by $H$ as $H$ is constrained to vanish. 
In the Lagrangian formulation the corresponding gauge transformations are Yang-Mills type gauge transformations and spacetime 
diffeomorphisms. \\
\\
To arrive at a predictive theory, we may employ the reduced phase space formalism. We outline here 
its simplest version, see \cite{3c,3d} for the general case. First the abstract description: The 
constraint surface $\bar{\Gamma}$ is the submanifold of $\Gamma$ where $Z_I=C_I=0$ holds 
for all $I$. By the first class property, the Hamiltonian flow of the constraints 
$Z_I,C_I$ is tangential to $\bar{\Gamma}$. This means that we can classify points 
$(Q,P,X,Y)$ on $\bar{\Gamma}$ into gauge orbits $[(K,M,X,Y)]$, that is, gauge 
equivalence classes. Here $(K,M,X,Y)$ and $(K',M',X',Y')$ on $\bar{\Gamma}$  
are equivalent iff there exist $f^I,g^I$ such that the Hamiltonian flow of $f^I Z_I+g^I C_I$ 
maps $(K,M,X,Y)$ to $(K',M',X',Y')$. The set of those gauge equivalence classes defines 
the abstract reduced phase space $\hat{\Gamma}$. It is a universal object 
and contains the full gauge invariant information about the system but it is of little 
practical interest because it does not tell us how to make contact with observation.  

Now for the practical description. We exploit the fact that in the present situation 
the phase space is a Cartesian product $\Gamma_p \times \Gamma_s$ where the labels 
p,s denote primary and secondary respectively. The constraints $Z,C$ respectively act non-trivially 
on $X,Y$ and $K,M$ only hence the reduced phase space construction can be performed factor 
wise with respect to $Z,C$ respectively. We consider first the secondary factor and  
split the canonical pairs
$(K^A, M_A)$ into two disjoint sets $(q^a,p_a),\;(x^I,y_I)$ where the number of pairs in the 
second entry matches the number of $C_I$. This split is in principle arbitrary, it is guided 
by the practical requirement that the purpose of this split is to solve $C_I=0$ for 
\be \label{2.2}
\bar{C}_I:=y_I+h_I(x;q,p)=0  
\ee
at least locally in $\Gamma_s$. Here locality refers to the typical situation that 
$C_I$ depends quadratically on $y$ so that one may need to solve square roots leading to 
branches. We assume that such a branch has been chosen and denote it by $\bar{\Gamma}_s$.
By a well known argument we have $\{\bar{C}_I,\bar{C}_J\}=0$ on that branch. 
The constructions reviewed below are restricted to one branch. Information on how these 
branches are assembled will be given in section \ref{s7}.

A gauge cut of the secondary constraint surface is a submanifold of $\bar{\Gamma}_s$ that intersects the 
gauge orbits everywhere transversally. We consider a one parameter set of 
gauge fixing conditions 
\be \label{2.3}
G^I(t):=x^I-k^I(t)
\ee 
which depend explicitly and non-trivially on the time parameter $t$. Note that 
$\{G^I(t),G^J(t'))\}=0,\; \{\bar{C}_I,G^J(t)\}=\delta_I^J$.
It is required that the subset $\hat{\Gamma}_{s,t},$ of $\bar{\Gamma}_s$ where 
(\ref{2.3}) holds is a a gauge cut of the constraint surface for each $t$ such that 
different cuts do not intersect. This holds at least locally in phase space as by assumption we
have $C_I=M_I^J \bar{C}_J$ with non-singular $M$. Then the gauge cuts foliate the constraint surface and 
the gauge cuts are all gauge equivalent, that is, for any two $t_1,t_2$ we find $g^I$ depending 
on $t_1,t_2$ such that the flow of $g^I C_I$ maps between the cuts defined by $G^I(t_1)=0,G^I(t_2)=0$. 
In this situation, the gauge cuts can be identified under gauge transformations.
The free canonical coordinates of each $\hat{\Gamma}_{s,t}$ are the $(q^a,p_a)$ referred to a the true 
degrees of freedom while the fixed respectively constrained degrees of freedom $(x^I,y_I)$ are 
referred to as the gauge degrees of freedom. The canonical pairs $q,p$ can therefore be 
considered as coordinates of a representative or concrete realisation 
of the abstract reduced phase space $\hat{\Gamma}_s$.

We now introduce the reduced Hamiltonian $h_s(t)$ which depends in general explicitly on $t$
depending on the form of the functions $k^I(t)$. We construct the functions 
$X^I_\ast(q,p;t)$ by asking that the equations 
\be \label{2.4}
\frac{d G^I(t)}{dt}:=\{H,G^I(t)\}+ \frac{\partial G^I(t)}{\partial t}
=X^J \;\{C_J,x^I\}-\dot{k}^I(t)=0
\ee 
hold when $C=G=X-X_\ast=0$. This linear equation has the explicit solution 
$X^I_\ast=\dot{k}^J\;[(M^{-1})_J^I]_{C=G=0}$. Condition (\ref{2.4}), referred to as stability condition, makes sure that 
the gauge cut is preserved in time $t$ when $H$ is interpreted as generating time translations in $t$.
The reduced Hamiltonian is the effective Hamiltonian depending only on the true degrees of freedom $(q,p)$ such 
that the equations of motion on functions $F$ depending only on $(q,p)$ agree with those of $H$ when    
$C=G=X-X_\ast=0$
\be \label{2.5}
\{h_s(t),F\}=\{H,F\}_{G=C=X-X_\ast=0}
\ee
This has the explicit solution 
\be \label{2.6}
h_s(q,p;t)=\dot{k}^I(t)\; h_I(x=k(t);q,p)
\ee

Finally we reduce the other factor of the kinematical phase by asking that also 
the condition $X-X_\ast=0$ be stabilised 
\be \label{2.7}
\frac{d (X^I-X^I_\ast(t))}{dt}:=\{H,X^I-X^I_\ast(t)\}+ \frac{\partial (X^I-X^I_\ast(t))}{\partial t}
=V^I -\frac{d X^I_\ast(t)}{dt}=0
\ee 
which fixes also the undetermined velocity $V^I$ of $X^I$ as that of $X^I_\ast$. 

Thus a 1-parameter family of gauge fixing conditions induces a concrete reduced phase space representative 
in terms of true degrees of freedom $(q,p)$ that comes with its own, non-vanishing Hamiltonian $h_s(q,p;t)$. 
The procedure therefore restores predictability and describes the dynamics of all gauge 
fields $X=X_\ast,Y=0,x=k,y=-h$ in terms of the dynamics of the true fields $q,p$. In the language of reference 
frames, the fields $x$ are called reference fields and together with the gauge fixing condition $G=x-k(t)$ define 
a reference frame.  

In the context of General Relativity and when considering geometrical clocks 
by imposing gauge conditions on components of the metric, we find that all metric components 
are expressed in terms of the true degrees of freedom which in this case consists of two polarisation 
degrees of freedom. In the Lagrangian language, conditions of the kind (\ref{2.3}) are typically 
imposed in order to fix residual gauge freedom that an incomplete gauge fixing condition like 
the de Donder gauge which is of the type (\ref{2.7}) is not able to fix, see 
\cite{8} for more information on the relation between Lagrangian and Hamiltonian treatment of gauge 
invariance.\\
\\
The only task left to do is to justify why we identify the true degrees of freedom $q,p$ with observables. 
After all, $q,p$ do not Poisson 
commute with at least the constraints $C$, they are not gauge invariant and thus not 
assigned the status of an observable in the mathematical sense. We now construct the gauge invariant relational observables and a physical 
Hamiltonian that drives their dynamics explicitly and show that there exists a canonical transformation 
to the true degrees of freedom and reduced Hamiltonian respectively. The transformation is 
defined on smooth functions $F$ of the secondary space $\Gamma_s$. It is given by the so called 
observable map
\be \label{2.8}   
O_F(t):=[e^{g^I W_I} \dot F]_{g=-G(t)}
\ee
where $W_I=\{\bar{C}_I,.\}$ denotes the Hamiltonian vector field of $\bar{C}_I$. Note that one first 
needs to compute the Hamiltonian flow of $g^I W_I$ with $g$ considered as phase space 
independent parameters before equating it with the phase space dependent function $-G(t)$.

Let 
\be \label{2.9}
\{F_1,F_2\}_\ast:=\{F_1,F_2\}+\{F_1,\bar{C}_I\}\;\{G^I(t),F_2\}-\{F_2,\bar{C}_I\}\;\{G^I(t),F_1\}
\ee 
be the Dirac bracket subordinate to $G(t)$. Note that it is in fact independent of $t$. 
We call an equation to hold weakly when it holds strictly on the constraint surface $Z=\bar{C}=0$
and strongly when it holds even away from it.  
We now list, without proof, a few properties of the map (\ref{2.8}) which hold weakly for any fixed number $t$ for real valued constraints and 
gauge fixing conditions, see e.g. \cite{3c,3d} for all proofs:\\  
1. $O_\cdot(t)$ is gauge invariant:  $\{\bar{C}_I,O_F(t)\}=0$ and $\{Z_I,O_F(t)\}=\{\bar{C}_I,O_F(t)\}=0$ 
even strongly.\\
2. $O_\cdot(t)$ is a $^\ast-$algebra homomorphism: $O_{F_1+F_2}(t)=O_{F_1}(t)+O_{F_2}(t),\;
O_{F_1\cdot F_2}(t)=O_{F_1}(t)\cdot O_{F_2}(t),\;O_{F^\ast}(t)=[O_F(t)]^\ast$.\\
3. $O_\cdot(t)$ acts trivially on the secondary reference fields $O_{X^I}(t)=k^I(t)\cdot 1, \;
O_{Y_I}(t)=-O_{h_I}(t)$.\\
4. $O_\cdot(t)$ is a Dirac bracket homomorphism: $\{O_{F_1}(t),O_{F_2}(t)\}_\ast=O_{\{F_1,F_2\}_\ast}(t)$.\\
Moreover, restricted to functions 
$F\in C^\infty(\hat{\Gamma}_s)$ on the representative of the reduced phase space
defined by the true degrees of freedom, not depending explicitly on 
$t$, we have strongly:\\
5. $O_\cdot(t)$ is a canonical transformation: $\{O_{F_1}(t),O_{F_2}(t)\}=O_{\{F_1,F_2\}}(t)$.\\
6. $t\mapsto O_\cdot(t)$ has $O_{h_s(t)}(0)$ as generator:  $\frac{d O_F(t)}{dt}=\{O_{h_s(t)}(t),O_F(t)\}$.

It is property 5. which justifies the identification of the true degrees of freedom $(q,p)$ 
with the relational observables at any fixed time $t=t_0$, say $t_0=0$, i.e. $Q^a:=O_{q^a}(0), P_a:=O_{p_a}(0)$.
In particular $\{P_a,Q^b\}=\{p_a,q^b\}=\delta_a^b$ and $\{Q^a,Q^b\}=\{q^a,q^b\}=0$ and 
$\{P_a,P_b\}=\{p_a,p_b\}=0$. 
Statement 6. means that $O_F(t)$ is the time evolution of $O_F(0)$ with respect to 
the physical Hamiltonian 
\be \label{2.9a}
H_s(t)=h_s(t; q\to O_q(0), p\to O_p(0))
\ee 
that is 
\be \label{2.10}
O_F(t)={\cal P}_r(\exp(int_0^t\; dr\; \{H_s(r), .\})\cdot O_F(0))
\ee
where the path ordering symbol orders the latest time dependence to the right.
We verify by checking that (\ref{2.10}) solves 6. and clearly the initial value of $O_F(t)$ at 
$t=0$ is $O_F(0)$ which is reproduced by (\ref{2.10}). We have, using (\ref{2.10}) 
\ba \label{2.11}
\frac{d O_F(t)}{dt} &=& {\cal P}_r(\exp(_0^t\; dr\; \{H_s(r), .\})\cdot \{H_s(t),O_F\}
={\cal P}_r(\exp(\int_0^t\; dr\; \{H_s(r), .\})\cdot O_{\{h_s(t),F\}}(0)
\nonumber\\
&=& \{h_s(t),F\}(0)]({\cal P}_r(\exp\int(_0^t\; dr\; \{H_s(r), .\})\cdot O_q(0),
{\cal P}_r(\exp(\int_0^t\; dr\; \{H_s(r), .\})\cdot O_p(0)](q,p)
\ea
by elementary properties of canonical transformations and we understand $q,p$ as the coordinate 
functions $q(q_0,p_0)=q_0,p(q_0,p_0)=p_0$ that return the corresponding entry at $(q_0,p_0)$.
On the other hand, 6. can be written
\be \label{2.12}
\frac{d O_F(t)}{dt}=\{O_{h_s(t)}(t),O_F(t)\}=O_{\{h_s(t),F\}}(t)=
\{h_s(t),F\}(O_q(t),O_p(t))
\ee
by properties of the map $O_\cdot(t)$. Thus both equations satisfy the same ordinary
differential equation iff (\ref{2.10}) holds for the coordinate functions $F=q, F=p$. 
But choosing $F=q$ or $F=p$ in (\ref{2.11}) and (\ref{2.12}) we obtain 
for $z(t):=(O_q(t)),O_p(t)),\; z'(t)=({\cal P}_r(\exp(_0^t\; dr\; \{H_s(r), .\})\cdot O_q(0),
{\cal P}_r(\exp(_0^t\; dr\; \{H_s(r), .\})\cdot O_p(0))$ and $v_t=(\{h_t,q\},\{h_t,p\})$ 
the equations $\dot{z}(t)=v_t(z(t)),\; \dot{z}´(t)=v_t(z'(t)), \; z(0)=z'(0)$. Thus 
by existence and uniqueness theorems, $z(t)=z'(t)$.

Under the canonical transformation $(Q,P)\to (q,p)$ the physical Hamiltonian becomes the 
the reduced Hamiltonian 
\be \label{2.13}
h_s(t)=\dot{k}^I(t) h_I(x=k(t),q.p)
\ee
The crucial point of this canonical transformation is now the following: When 
dealing with only one split of degrees of freedom into gauge and true ones, we do not need to bother 
working out the infinite series $\sum_{n=0}^\infty \frac{1}{n!} ([g^ I W_I]^n\cdot F)_{g=-G(0)}$ 
because we only need to deal with the $Q,P$ and their Poisson brackets among themselves and with  
$H_s(t)$, both classically and in quantum theory where $Q,P, H_s(t)$ are canonically quantised. 
Mathematically, this is completely equivalent to working with $q,p,h_s(t)$ which is more convenient 
because we never need to mention the complicated maps $O_\cdot(t)$. However, when working with
$q,p,h_s(t)$ we must remember that these really are to be physically interpreted as the
gauge invariant and therefore observable relational observables $Q,P,H_s(t)$. Moreover, 
the explicit form of the observable maps $O_\cdot(t)$ becomes important when switching reference frames.

\section{Relational reference frame transformation}
\label{s3} 
 
In the previous section we have reviewed how a typical gauge system within our class of interest 
can be described in terms 
of gauge invariant relational observables. This requires by definition as an input a choice of reference fields 
and a choice of gauge fixing conditions on these reference fields that we called gauge degrees of freedom. 
The choice of reference fields and gauge fixing conditions on them defines a relational reference frame in this 
context. Since 
this is a central notion for the present work and to distinguish it from other possible 
notions of reference frame, we make it a definition.
\begin{Definition} \label{def3.1} ~\\
i. \\
Given a gauge system of the class described in section \ref{s2}, a {\sf relational reference frame (RRF)} is 
a pair $(x,k(.))$ consisting of 1. a choice of reference fields $x^I$ in the list 
$(K^A,M_A)=((q^a,p_a),(x^I,y_I))$ of canonical pairs  
equal in number to the 
number of secondary first class constraints $C_I$ such that that the $C_I$ can be solved locally 
for the conjugate momenta $y_I$ and 2. a choice of 1-parameter system of 
gauge fixing conditions $t\mapsto G^I(t)=x^I-k^I(t)=0$ where 
$k^I(t)$ are certain constant functions on phase space and $\dot{k}^I(t)\not=0$ for at least one 
index $I$.\\
ii.\\
A change of reference frame $(x,k(.))\mapsto (\hat{x},\hat{k}(.))$ is either a change of 
choice of both reference field and gauge fixing condition $x\not=\hat{x}, \;k\not=\hat{k}$ or 
a change of choice of just the the gauge fixing condition for the same choice of reference fields
$x=\hat{x}, \;k\not=\hat{k}$. \\
iii.\\
For a pair $(t,\hat{t})$ of corresponding parameters, we call a canonical transformation 
between the corresponding relational observables, or equivalently true degrees of freedom, 
a {\sf relational reference frame transformation (RRFT)} at $t,\hat{t}$.
\end{Definition}
Since the choice of reference frame is rather  
arbitrary it is a valid question how the physical description depends on the choice of reference 
frame in particular because the abstract reduced phase space defined in terms of gauge orbits does 
not require this extra structure. Accordingly, in this section we start with the same kinematical setup as before but 
now introduce two different splits of the secondary phase space $\Gamma_s$ with coordinates 
$(K^A,M_A)$ into different pairs $(q^a,p_a),(x^I,y_I)$ and $(\hat{q}^a,\hat{p}_a),(\hat{x}^I, \hat{y}_I)$ 
respectively. We do not consider the most general situation but restrict attention to the following:
Let $a,b,c,..\in {\cal A},\;I,J,,..\in {\cal I}$. We consider disjoint unions ${\cal A}={\cal A}_1\cup {\cal A}_2$ and
${\cal I}={\cal I}_1\cup {\cal I}_2$ with $|{\cal A}_2|=|{\cal I}_2|$ such that 
\be \label{3.1}
q^a=\hat{q}^a;\;a\in {\cal A}_1,\;x^I=\hat{x}^I;\;I\in {\cal I}_1,\;q^a=\delta^a_I\hat{x}^I, \;x^I=\delta^I_a \hat{q}^a;\;a\in {\cal A}_2,\;I\in {\cal I}_2
\ee
and similar for $p,y,\hat{p},\hat{y}$.
Thus partially the designated true and gauge degrees of freedom coincide but partially the designated true degrees of freedom 
of the first split are gauge degrees of freedom in the second split and vice versa. We also allow ${\cal A}_2={\cal I}_2=\emptyset$
which describes the change of just the choice of gauge fixing condition. The generalisation of this particular choice of 
change between reference fields would be an arbitrary canonical transformation on the kinematical phase space between 
$K^A,M_A$ and $\hat{K}^A,\hat{M}_A$ and then performing the same split of the index set $A\in T=T_1\cup T_2,‑ T_1\cap T_2=
\emptyset$ to designate $a=A, A\in T_1; I=A,\; A\in T_2$.    

We now carry out the construction of the previous section for both splits and have the pairs of 
solved for constraints and gauge fixing conditions
$\bar{C}_I=y_I+h_I(x;q,p),\;G^I=x^I-k^(t)$ and 
$\hat{\bar{C}}_I=\hat{y}_I+\hat{h}_I(\hat{x};\hat{q},\hat{p}),\;\hat{G}^I=\hat{x}^I-\hat{k}^(\hat{t})$
respectively.
It is of course required that the branches that the two sets of solved for constraints select are 
identical, that is, $\bar{C}=0\;\;\Leftrightarrow\;\; \hat{\bar{C}}=0$ otherwise we reduce different 
sectors of the phase space and one would not expect any obvious relation between the resulting 
reductions. Likewise we obtain reduced Hamiltonians 
$h_s(q,p;t)=\dot{k}^I(t)\;h_I(x=k(t);q,p)$ and 
$\hat{h}_s(\hat{q},\hat{p};t)=\dot{\hat{k}}^I(t)\;\hat{h}_I(\hat{x}=\hat{k}(t);\hat{q},\hat{p})$
respectively. 

The question asked in the introduction is in what sense these two reductions are equivalent?
In particular, is there a canonical transformation or symplectomorphism 
$S_{t,\hat{t}}:\;(q,p)\mapsto (\hat{q,\hat{p}}):=
\hat{S}(q,p)$ and a reparametrisation $T:\; t\mapsto \hat{t}=T(t) $ such that $h_s(q,p;t)=\hat{h}_s(\hat{k}(T(t));S_{t,T(t)}(q,p))$?
We expect a reparametrisation that accomplishes the identification of Hamiltonians under this transformation
to exist only in the case that only one of the functions $k^I(t)$ and one 
of the functions $\hat{k}^I(t)$ display non-trivial time dependence, otherwise one would need to 
deal with what is called {\it multi-fingered} time in the literature, i.e. one would need as many time parameters 
as there are non-trivial functions $k^I,\hat{k}^I$ i.e. in the worst case as many as there are 
secondary constraints. For instance 
in General Relativity in four spacetime dimensions one may consider four functions $\phi^\mu,\;\mu=0,..3$
which play the role of the $x^I$ and imposes gauge conditions of the form $\phi^0=k(t),\; \phi^a=\varphi^a(z),\; a=1,2,3$ where $(t,z^a)$
are spacetime coordinates such that $(t,z)\mapsto (k(t),\varphi(z))$ is a spacetime diffeomorphism. Then only the 
zero mode with respect to $\partial_{z^a}$ and only of $\phi^0$ displays a non-trivial time dependence.
However, we will not limit ourselves to such choices of $k^I(t)$ in what follows because it will turn out 
that while natural canonical transformations between different reference frames exist, generically the reduced 
Hamiltonians are not identical under the corresponding pull-back for very good reasons.
       
Indeed, there are two obvious candidates at least for the canonical transformation. The first is the 
{\it identity reference frame transformation (IRFT)}
\be \label{3.2a}
(\hat{q,\hat{p}}):S_{t,\hat{t}}(q,p):=(q,p)
\ee
which in fact does not depend on $t,\hat{t}$.
This can be extended as canonical transformation to the full secondary phase space by (\ref{3.1}) and amounts 
to partly switching between what are called gauge and true degrees of freedom respectively in the 
two reductions without any dynamical input. It is entirely kinematical in nature. Under the identification 
of $(q,p)\to (O_q(t),O_p(t))$ and $(\hat{q},\hat{p})\to (\hat{O}_{\hat{q}}(\hat{t}),\hat{O}_{\hat{p}}(\hat{t}))$ 
we may also use it as a canonical transformation 
at the level of relational Dirac observables of the two reductions at times $t,\hat{t}$ respectively 

The second is based on the fact that relational observables are weakly invariant under the flow of 
the constraints. Denote by $O_\cdot(t)$ and $\hat{O}_\cdot(\hat{t})$ respectively the maps introduced 
in the previous section that map any smooth functions $F$ on the secondary phase space to the 
corresponding relational observable subordinate to $G(t)$ and $\hat{G}(\hat{t})$ respectively. Note that 
$\{\bar{C},O_F(t)\}=\{\hat{\bar{C}},\hat{O}_F(t)\}=0$ strongly but 
$\{\hat{\bar{C}},O_F(t)\}=\{\bar{C},\hat{O}_F(t)\}=0$ only weakly. Then we have the weak 
identity
\be \label{3.2} 
\hat{O}_F(\hat{t})=O_{\hat{O}_F(\hat{t})}(t)
\ee
for any $t,\hat{t}$. The left hand side is a relational observable with respect to the second 
reduction. It is a certain invariant function $f(K,M)$ on the full secondary phase space, namely 
the image of $F(K,M)$ with respect to the second observable map $\hat{O}_\cdot(\hat{t})$ which 
can be used as an input for the first observable map $O_\cdot(t)$ and therefore results 
in a relational observable with respect to the first reduction. Fix $t,\hat{t}$ and 
define 
\be \label{3.3}
\hat{Q}^a:=\hat{O}_{\hat{q}^a}(\hat{t}),\;
\hat{P}_a:=\hat{O}_{\hat{p}_a}(\hat{t}),\;
Q^a:=O_{q^a}(t),\;
P_a:=\hat{O}_{p_a}(t),\;
\ee
We may use $\hat{Q},\hat{P}_a$ or alternatively $Q^a,P_a$ as elementary Dirac observables
separating the points of the abstract reduced phase space.
The weak identity (\ref{3.2}) defines a weak canonical transformation between them since
\be \label{3.4}
\{\hat{P}_a,\hat{Q}^b\}=\delta_a^b=\{O_{\hat{P}_a}(t),O_{\hat{Q}^b}(t)\}
=\{O_{\hat{P}_a}(t),O_{\hat{Q}^b}(t)\}_\ast=O_{\{\hat{P}_a,\hat{Q}^b\}_\ast}(t)
=O_{\{\hat{P}_a,\hat{Q}^b\}}(t)=\delta_a^b
\ee
indeed. Here $\{.,.\}_\ast,\;\{.,.\}_{\hat{\ast}}$ denote the Dirac brackets 
subordinate to the first and second reduction respectively and it was used that the 
Dirac bracket between two Dirac observables weakly equals the Poisson bracket. Suppose 
that we have worked out the explicit expressions
\be \label{3.5}
\hat{Q}^a=f^a(q,p,x,y;\hat{t}),\;
\hat{P}_a=g_a(q,p,x,y;\hat{t})
\ee
on the kinematical phase space. The right hand side of (\ref{3.5}) can actually be written just 
in terms of $\hat{q},\hat{p},hat{x}$ at any point of the kinematical phase space 
but we then use (\ref{3.1}) to write it in terms 
of $q,p,x,y$. Then we find weakly the identity
\be \label{3.5a}
\hat{Q}^a=f^a(Q,P,k(t),-h(k(t);Q,P);\hat{t}),\;
\hat{P}_a=g_a(Q,P,k(t),-h(k(t);Q,P);\hat{t})
\ee
using all the properties of the observable map listed in the previous section. Note that 
(\ref{3.5a}) is an identity only on the constraint surface. We now take the right hand side 
of (\ref{3.5a}) as a {\it definition} for the map $S_{t,\hat{t}}(Q,P)$ which 
is now a {\it strong} canonical transformation because we have used the constraints $C=0$ 
to express $y=-h(x,q,p)$ and (\ref{3.4}) is an identity when using the constraints.

By the same method, we also find the inverse of $S_{t,\hat{t}}$, denoted $S_{\hat{t},t}$ by 
exchanging the roles of the reference frames in (\ref{3.2}). This is because we have 
the weak identity 
\be \label{3.5b}
\hat{O}_F(\hat{t})=O_{\hat{O}_F(\hat{t})}(t)=\hat{O}_{O_{\hat{O}_F(\hat{t})}(t)}(\hat{t})
\ee
which we can turn into a strong one using the constraints.\\ 
\\
\\ 
This 
establishes the canonical transformation concretely at the level of the relational 
observables. It expresses the fact that the number of algebraically independent 
Dirac observables of course is independent of the relational scheme in which one 
finds them. We can now use again the fact that the maps $O_\cdot(t),\;\hat{O}_\cdot(\hat{t})$
are canonical transformations between true degrees of freedom and relational Dirac 
observables respectively at any times $t,\hat{t}$. This enables us to write the canonical transformation 
directly at the level of true degrees of freedom
\be \label{3.6}
\hat{q}^a=f^a(q,p,k(t),-h(k(t);q,p);\hat{t}),\;
\hat{p}_a=g_a(q,p,k(t),-h(k(t);q,p);\hat{t})
\ee
which defines the same transformation $S_{t,\hat{t}}$ as above just that 
the relational observables were replaced by the corresponding true degrees of freedom. 
We call (\ref{3.5a}) or equivalently (\ref{3.6}) the {\it relational reference frame transformation} (RRFT).
It is also canonical transformation but in contrast to the identity reference frame transformation 
it encodes dynamical information.

The RRFSM $S_{t,\hat{t}}$ has the following 
interpretation, in terms of the true degrees of freedom: The map $O_\cdot(t)$ has the property 
that it installs the gauge $x=k(t)$ from anywhere in the full phase space $\Gamma_s$ and in particular 
on the constraint surface $\bar{\Gamma}_s$. It therefore "projects" $\bar{\Gamma}_s$ to 
$\hat{\Gamma}_{s,t}$ in the sense of sets or smooth functions on $\bar{\Gamma}_s$ to smooth 
functions on $\hat{\Gamma}_{s,t}$ by pullback. It is therefore not even a bijection. The same 
holds for $\hat{O}_\cdot(\hat{t})$. The way this is accomplished is as follows: 
In order to fully understand the mechanism, we must distinguish between the phase space 
coordinate functions $q,p,x,y$ and the values they take on a phase space point $z=(z^1,z_1,z^2,z_2)$, 
specifically $q(z)=z^1$ etc.
The constraint
$\bar{C}=y+h(x;q,p)$ has the property $\{\bar{C},x\}=1$. Therefore, the gauge transformation 
that maps the value $z^2$ of the phase space function $x$ to the value $\tilde{z}^2$ is given by the 
canonical transformation $\exp((\tilde{z}^2-z^2)\{\bar{C},.\})$ as the phase space function $x$ is 
mapped to the new phase space function  
$\tilde{x}=\exp((\tilde{z}^2-z^2)\{\bar{C},.\})\cdot x=(x+\tilde{z}^2-z^2)\;1$
where $1$ is the constant function equal to unity on the phase space. Evaluating the 
new function $\tilde{x}$ at $z$ yields $\tilde{x}(z)=\tilde{z}^2$ provided that $x(z)=z^2$. But this works 
only from the point $z$. If we want to reach the value $\tilde{z}^2$ from anywhere, then we must 
use a different map, namely $[\exp((\tilde{z}^2-z^2)\{\bar{C},.\})]_{z^2\to x}$ which consists 
of concatenation of the previous canonical transformation with a substitution map which consists 
in replacing the numerical value $z^2$ by the phase space function $x$. But this is exactly 
the definition of $O_\cdot(t)$ with $k(t)=\tilde{z}^2$. Now we get the 
new phase space function $\tilde{x}=O_x(t) = \tilde{z}^2\;1$ which is constant on the phase space,
hence evaluating it at any point $z$ indeed yields $\tilde{z}^2$. Thus $O_\cdot(t)$ installs the 
value $k(t)$ of $x$ from anywhere on $\Gamma_s$. Similar considerations hold for $\hat{O}_\cdot(\hat{t})$. 

Thus, the map $\hat{O}_\cdot(\hat{t})$ gives us the possibility to map the entire 
$\hat{\Gamma}_{s,t}$ to the entire $\widehat{\hat{\Gamma}}_{s,\widehat{t}}$. Hence
we consider the phase space functions $q,p,x,y$ and 1. map them with $\hat{O}_\cdot(\hat{t})$ and then
2. evaluate these at points $z\in \hat{\Gamma}_{s,t}$. The first step yields on the full secondary phase space 
$(q,p,x,y)\mapsto (\hat{k}(\hat{t})\;1,\hat{O}_p(\hat{t}),\hat{O}_x(\hat{t}),\hat{O}_y(\hat{t}))$.
On the constraint surface, this information is redundant because $p=-\hat{h}(q;x,y)$ so we may 
drop the entry $ \hat{O}_p(\hat{t})$ and we may drop the entry $\hat{k}(\hat{t})\;1$ since we know 
that we are on $\widehat{\hat{\Gamma}}_{s,\hat{t}}$. This leaves us with the phase space
functions $(\hat{O}_x(\hat{t}),\hat{O}_y(\hat{t})$ as entries. In the second step 
we restrict these to points 
on $\hat{\Gamma}_{s,t}$. This means that the functions $(\hat{O}_x(\hat{t}),\hat{O}_y(\hat{t})$ 
when written in terms of $q,p,x$ using $y=-h(x;q,p)$ can be replaced by the same functions just 
that $x$ is replaced by $k(t)$. It follows 
\be \label{3.6a}
S_{t,\hat{t}}(q,p)=(\hat{O}_x(\hat{t}),\hat{O}_y(\hat{t})_{y=-h(x,q,p),x=k(t)}
\ee
at the level of true degrees of freedom
But this is exactly the same as $(O(t)_{\hat{O}_x(\hat{t})}(t),\; O(t)_{\hat{O}_y(\hat{t})}(t))$, i.e. 
(\ref{3.5a})  
at the level of relational observables, just that $q,p$ are replaced by $O_q(t), O_p(t)$, i.e.
(\ref{3.6})\\  
\\
To interpret the dynamics generated by the reduced Hamiltonian on the reduced phase space 
geometrically, we consider 
a gauge orbit $\gamma\in \hat{\Gamma}_s$ on the constraint surface $\bar{\Gamma}_s$. We use the notation 
$[q_0,p_0,x_0]$ to denote the gauge equivalence class of $(q_0,p_0,x_0,y_0=-h(x_0;q_0,p_0))$.
We will drop the $y_0$ entry for what follows as it is determined by the others. Given $\gamma$ we 
consider its intersection with the gauge cuts $\hat{\Gamma}_{s,t}=\bar{\Gamma}_{s|x=k(t)}$ 
and denote it as $(q_t(\gamma),p_t(\gamma),k(t))$. Note that we distinguish again between 
the phase space coordinate functions $q,p,x,y$ and points $(q_0,p_0,x_0,y_0)$ where 
$q(q_0,p_0.x_0,y_0)=q_0$ etc. The Dirac observable $O_f(t)$ is a function on $\bar{\Gamma}_s$ 
for a function $f$ of $q,p$ which 
is a constant on each gauge orbit $\gamma$, i.e. $O_f(t)(q_0,p_0,x_0)$ depends only on $[q_0,p_0,x_0]$.
Specifically, for our given gauge orbit $O_q(t)(q_t(\gamma),p_t(\gamma),k(t))=q_t(\gamma)$ and 
$O_p(t)(q_t(\gamma),p_t(\gamma),k(t))=p_t(\gamma)$. Consider now $t'=t+\delta t$. 
Since $[q_t(\gamma),p_t(\gamma),k(t)]=[q_{t'}(\gamma),p_{t'}(\gamma),k(t')]$ we have the 
identity 
\ba\label{3.6b}
&& O_{q^a}(t')(q_{t'}(\gamma),p_{t'}(\gamma),k(t')) = q^a_{t'}(\gamma)=O_{q^a}(t')(q_t(\gamma),p_t(\gamma),k(t))
\nonumber\\
&=& [q^a+(k^I(t')-k^I(t))\{h_I,q^a\}+O(\delta t^2)](q_t(\gamma),p_t(\gamma),k(t))
\nonumber\\
&=& q^a_t(\gamma)+[\delta t]\;\{\dot{k}^I(t) h_I(k(t;.,.),q^a\}(q_t(\gamma),p_t(\gamma)) +O(\delta t ^2)
\ea
where we used the explicit series expression for the relational observables, thus
\be \label{3.6c}
q_{t+\delta}(\gamma)-q_t(\gamma)=[\delta t]\;\{\dot{k}^I(t) h_I(k(t;.,.),q^a\}(q_t(\gamma),p_t(\gamma)) 
\ee
to first order in $\delta t$ and similar for $p_a$. But the right hand side is precisely the infinitesimal dynamics generated 
by the reduced Hamiltonian $h_s(t;q,p)=\dot{k}^I(t) h_I(k(t); q,p)$ evaluated at $q_t(\gamma),p_t(\gamma)$
which can therefore be interpreted as the change of coordinates $(q_t(\gamma), p_t(\gamma),k(t))=\gamma\cap \hat{\Gamma}_{s,t}$ of the intersection
of a gauge orbit with the gauge cuts labelled by $t$ as $t$ varies. In this way while $\gamma$ is frozen, those coordinates 
change as we shift the cuts when $t$ varies. 

\section{Geometrical picture of the relational reference frame transformation and relational dynamics}
\label{s5} 

In this section we want to illustrate the relation between the abstract reduced phase space 
defined in terms of gauge orbits and the concrete reduced phase space in terms of gauge 
cuts and true degrees of freedom. This will enable a geometric understanding of the relational
reference frame 
transformation and relational dynamics. This geometrical picture emerges by evaluating the 
relational observables, which we have defined as functions on phase space, at points of the 
constraint surface.\\  
\\
It will be sufficient to consider a four dimensional kinematical secondary phase space
$\Gamma_s$
with coordinates $K^A, M_A;A=1,2$ and a single constraint $C$. Again we introduce the splits 
$q=K^1=\hat{x},p=M_1=\hat{y}, x=K^2=\hat{q}, y=M_2=\hat{p}$ and solve $C=0$ on a given branch 
$\bar{\Gamma}_s$ as 
$\bar{C}=y+h(x;q,p)$ and equivalently as  $\hat{\bar{C}}=\hat{y}+\hat{h}(\hat{x};\hat{q},\hat{p})$.

The branch is a three dimensional hypersurface in $\mathbb{R}^4$ and we can consider it graphically 
as (portion of) $\mathbb{R}^3$ by parametrising it by $K^1,K^2,M_1$ whose values we record 
on the $x^1,x^2,x^3$ axes of a Cartesian coordinate system. Then we obtain the embedding 
$\bar{\Gamma}_s\to \Gamma_s;\;(K^1,K^2,M_1)\to (K^1, K^2, M_1, M_2=-h(K^2;K_1,M_1))$. 
This (portion of) of $\mathbb{R}^3$ is filled by a congruence of gauge orbits. The 
gauge orbit through $(K^1,K^2,M_1)$ will be denoted by $\gamma=[(K^1,K^2,M_1)]$ and the set of gauge 
orbits defines the abstract reduced phase space $\hat{\Gamma}_s$. Given values $k(t),\hat{k}(\hat{t})$ 
we consider the two dimensional copies $\hat{\Gamma}_{s,t},\;\widehat{\hat{\Gamma}}_{s,\hat{t}}$ 
of (portions of) $\mathbb{R}^2$ defined by the gauge cut equations $G(t)=K^2-k(t)=0, \hat{G}(\hat{t})=
K^1-\hat{k}(\hat{t})=0$ respectively which in our diagramme are coordinate planes in the 
$K^1,M_1$ and $K^2,M_1$ direction respectively. As we vary $t,\hat{t}$ either family of planes foliates
$\bar{\Gamma}_s$. 

A gauge orbit $\gamma\in \hat{\Gamma}_s$, which is a 1-dimensional submanifold of $\bar{\Gamma}_s$,
is tangential to the 
the flow line of the Hamiltonian vector field of $\bar{C}$ or equivalently $\hat{\bar{C}}$. 
Points in $\hat{\Gamma}_{s,t}$ are coordinatised by the values of the phase space 
functions $q=K^1,p=M_1$ which coordinatise the phase space $\hat{\Gamma}_{s,{\sf true}}$  
and those in $\hat{\Gamma}_{s,\hat{t}}$ by the phase space functions 
$x=K^2,p=M_1$ or equivalently $x,y=M_2=-h(x;\hat{k}(\hat{t}),p)$ as we are on the constraint surface,
which coordinatise the phase space $\widehat{\hat{\Gamma}}_{s,{\sf true}}$.
Now as we have seen in section \ref{s3}, the map $S_{t,\hat{t}}$ that relates the functions $x,y$ on 
$\widehat{\hat{\Gamma}}_{s,\hat{t}}$ to the functions $q,p$ on $\hat{\Gamma}_{s,t}$ is defined
as the gauge transformation that installs the gauge $q=\hat{k}(\hat{t})$ from everywhere restricted 
to evaluation on points on $\hat{\Gamma}_{s,t}$. Thus we have the following geometrical or 
graphical interpretation of $S_{t,\hat{t}}$: Define the maps between abstract reduced phase space 
and gauge cuts by   
\be \label{5.1}
[.]:\;\bar{\Gamma}_s \to \hat{\Gamma}_s;\; (K^1,K^2,M_1)\mapsto [(K^1,K^2,M_1)],\;\;
\iota_t:\;\hat{\Gamma}_s\to \hat{\Gamma}_{s,t};\;\gamma\mapsto \gamma \cap \hat{\Gamma}_{s,t},\;\;
\hat{\iota}_{\hat{t}}:\;\hat{\Gamma}_s\to \widehat{\hat{\Gamma}}_{s,\hat{t}};\;\gamma\mapsto \gamma \cap 
\widehat{\hat{\Gamma}}_{s,\hat{t}}
\ee
and the canonical restriction and embedding maps  
\ba \label{5.2}
&&\pi:\bar{\Gamma}_s\to \hat{\Gamma}_{s,{\sf true}};\;(q,x,p)\mapsto (q,p),\;
\hat{\pi}:\bar{\Gamma}_s\to \widehat{\hat{\Gamma}}_{s,{\sf true}};\;(q,x,p)\mapsto (x,p),\;
\nonumber\\
&& \epsilon_t:\;\hat{\Gamma}_{s,{\sf true}}\to \hat{\Gamma}_{s,t};\;(q,p)\mapsto (q,k(t),p),\; 
\hat{\epsilon}_{\hat{t}}:\;\widehat{\hat{\Gamma}}_{s,{\sf true}}\to \widehat{\hat{\Gamma}}_{s,\hat{t}};\;(x,p)\mapsto (x,\hat{k}(\hat{t}),p) 
\ea
Then 
\be \label{5.3}
S_{t,\hat{t}}=\hat{\pi}\circ \hat{\iota}_{\hat{t}}\circ [.]\circ \epsilon_t
\ee
i.e. we take a point on $\hat{\Gamma}_{s,{\sf true}}$ emebed it into $\hat{\Gamma}_{s,t}$,
compute the gauge orbit through it,
record its intersection point with $\widehat{\hat{\Gamma}}_{s,\hat{t}}$ and restrict it to 
$\widehat{\hat{\Gamma}}_{s,{\sf true}}$.

Likewise, the graphical or 
geometrical interpretation of the dynamics determined by the reduced 
Hamiltonian $h_s(t)=\dot{k}(t)\;h(k(t),q,p$) as derived in section \ref{s3} is 
\be \label{5.4}
\alpha_{t,t'}=\pi\circ \iota_{t'}\circ [.]\circ \epsilon_{t}
\ee
i.e. we take a point on $\hat{\Gamma}_{s,{\sf true}}$ considered as a point on the initial $\hat{\Gamma}_{s,t}$, 
determine the gauge orbit through it, 
record its intersection point with the final $\hat{\Gamma}_{s,t'}$ and determine its coordinates 
considered as a copy of $\hat{\Gamma}_{s,{\sf true}}$. 
The dynamics of $q,p$ in $\hat{\Gamma}_{s,{\sf true}}$ is thus obtained as the trajectory 
swept out by the gauge orbit through the initial point with the intersections with the leaves of the 
foliation $t\mapsto \hat{\Gamma}_{s,t}$ of $\bar{\Gamma}_s$ as we change $t$.
This dynamical map has the analytical expression
$\alpha_{t,t''}={\cal P}_r\circ \exp(\int_t^{t'}\; dr\;\{h_s(r),\})$ where ${\cal P}_r$ orders time dependence 
from left (earliest) to right (latest) due the explicit time dependence.\\
\\
\\
This combines the 
frozen Dirac observable picture (the gauge orbits) with the dynamical relational observable or true degrees 
of freedom picture.

\begin{figure}
\includegraphics[width=20cm,height=16cm]{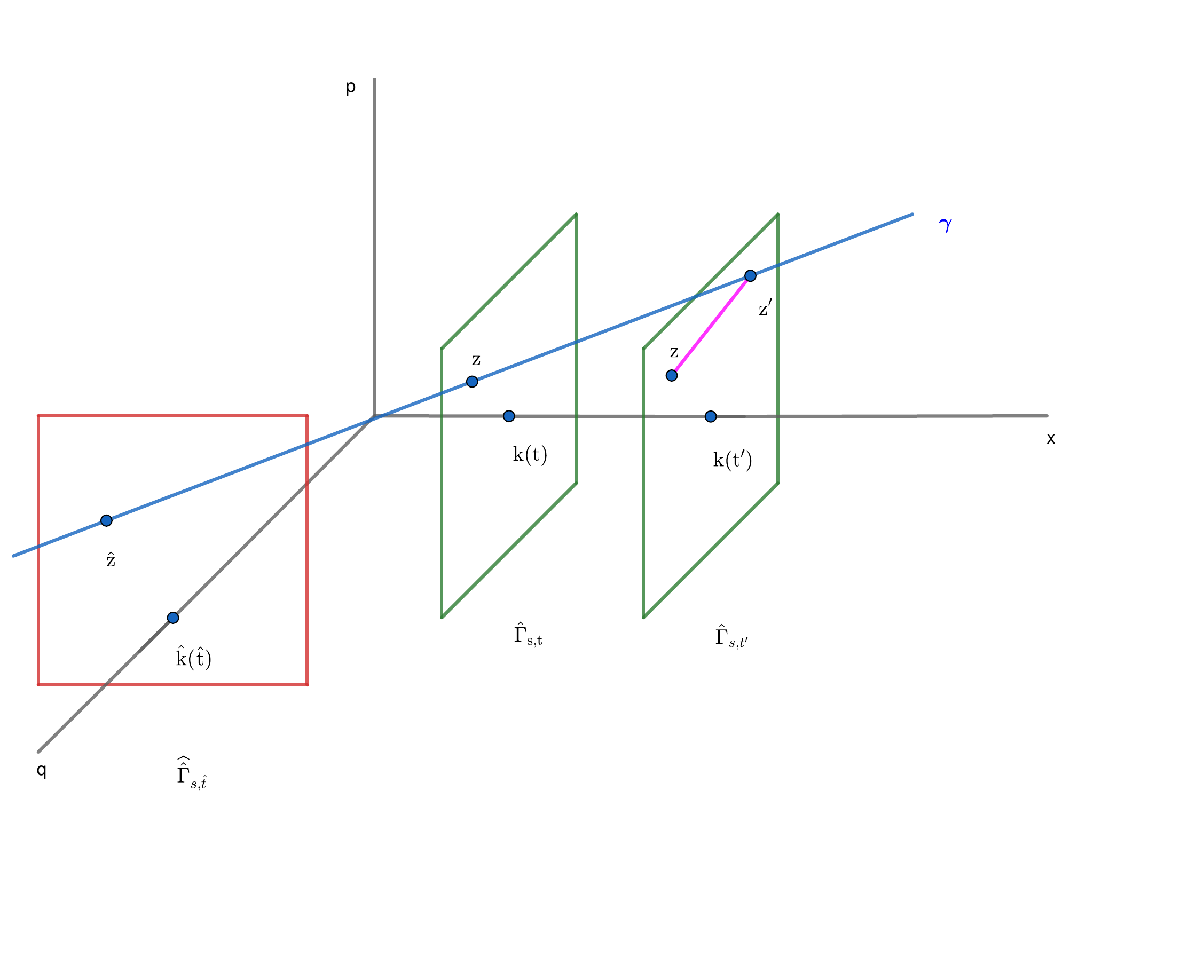}
\caption{
The branch of the constraint surface $\bar{\Gamma}_s$ defined 
by $y+h(x,p,q)=0\;\Leftrightarrow \; p+\hat{h}(q,x,y)$ of a 4-dimensional phase space
$\Gamma_s$. The $y$ direction is suppressed. 
It is filled by the congruence of gauge orbits $\gamma\in \hat{\Gamma}_s$ (blue) of the 
abstract reduced phase space. Displayed are two gauge cuts 
$\hat{\Gamma}_{s,t},\;\hat{\Gamma}_{s,t'}$ (green) corresponding to the reference frame $(x,k(.))$ and 
one gauge cut $\widehat{\hat{\Gamma}}_{s,\hat{t}}$ (red) corresponding to the reference frame $(q,\hat{k}(.))$.
All gauge orbits lie transversal to all gauge cuts of both foliations $t\mapsto \hat{\Gamma}_{s,t},\;
\hat{t}\mapsto \widehat{\hat{\Gamma}}_{s,\hat{t}}$ of $\bar{\Gamma}_s$.   
The two first gauge cuts are coordinatised by true degrees of freedom $(q,p)$, the latter is 
coordinatised by true degrees of freedom $(x,y=-h(x,\hat{k}(.),p)$. The relational reference 
frame transformation $\hat{z}=S_{t,\hat{t}}(z)$ is obtained by picking a point $z\in \hat{\Gamma}_{s,t}$,
determining the gauge orbit $\gamma$ on which it lies and determining its intersection point
$\hat{z}=\gamma\cap \widehat{\hat{\Gamma}}_{s,\hat{t}}$. The trajectory of $z\mapsto z'=\alpha_{t,t'}(z)$ (magenta) 
encoded by the 
$(x,k(.))$ reference frame is by determining the intersection point $z'=\gamma\cap \Gamma_{s,t'}$ of the same gauge orbit
$\gamma$ with 
a leaf of the same foliation at a later time $t'$ and projecting the segment of $\gamma$ between $z,z'$ into 
the $x=0$ plane.
}    
\label{fig5.1}
\end{figure}

\section{Images of reduced and physical Hamiltonians under relational reference frame transformation}
\label{s3a}

The question is now how the reduced or equivalently the physical Hamiltonians change under 
the identity or relational reference frame transformation.
In the following two subsections, we will present examples from simple 
(non-)relativistic mechanics in which neither 
the identity nor the relational reference frame transformation are such that 
the reduced Hamiltonian of the second reduction is the pull back of of the reduced 
Hamiltonian of the first reduction, not even up to a reparametrisation. These are examples in which both the choices of reference 
fields and gauge fixing conditions on them are changed. In the appendix we also consider 
a field theory example in which the reference fields are kept but the gauge fixing conditions 
are changed. Again the reduced or equivalently the physical Hamiltonians are not 
pullbacks under the relational reference frame transformation. We have chosen 
to defer the latter example to an appendix because it requires additional background material which would
distract from the main ideas of this article. In the final subsection we discuss the consequences and 
point out that as the explicit construction reveals, the failure of a match of reduced Hamiltonians after
pullback is not a loss of gauge invariance as either description is manifestly gauge invariant.

\subsection{Free relativistic particle}
\label{s3.1}

The free relativistic particle with rest mass $m$ in $D+1$ dimensional Minkowski space can be cast into the 
general framework of section \ref{s2} using the Hamiltonian
\be \label{3.7}
H=V\;Y+X\;C;\;C=M^A M_A + m^2
\ee
Indices are transported with the Minkowski metric and we dropped the index $I$ which here only takes 
one value.
It is obtained via partial Legendre transform from the Lagrangian 
\be \label{3.8}
L(K,U,X,V)=\frac{U^A U_A}{4 X}-m^2 X
\ee
which does not depend on $V$ and $K$ explicitly. Upon eliminating $X$ via its Euler Lagrange 
equation one obtains the more familiar reparametrisation invariant form 
\be \label{3.9}
L(Q,U)=-m\;\sqrt{-U^A U_A}
\ee 
for the positive root of $X$. Using (\ref{3.8}) we find primary constraint $Z=L_{,V}=0$ and 
$M_A=L_{,U^A}=U_A/(2X)$ so that $M_A U^A-L=X \; C$ indeed. Stability of the 
primary constraint yields the mass shell condition $C=0$ as secondary constraint. 

To reduce the system we decide on a branch of the constraint surface. We pick 
$M_0, M^1<0$ i.e. the particle moves to the future $U^0>0$ with negative velocity component 
$U^1<0$ along the $K^1$ axis. This enables to parametrise the trajectory alternatively 
in terms of $K^0$ or $K^1$. We consider the two reductions based on the splits
\ba \label{3.10}   
&& x=K^0,\;y=M_0;q^a:=K^a,\;p_a:=M_a;\;a=1,..,D
\nonumber\\
&& \hat{x}=K^1,\;\hat{y}=M_1;\hat{q}^1:=K^0,\;\hat{p}_1:=M_0,\;\hat{q}^a:=K^a,\;\hat{p}_a:=M_a;\;a=2,..,D
\ea 
and solve the secondary constraints as
\be \label{3.11}
\bar{C}=y+\sqrt{m^2+p_1^2+e}, \hat{\bar{C}}=\hat{y}+\sqrt{\hat{p}_1^2-m^2-\hat{e}};\;
e:=\sum_{a=2}^D p_a^2,\;
\hat{e}:=\sum_{a=2}^D \hat{p}_a^2
\ee
The corresponding gauge fixing conditions are $G(t)=x-k(t),\hat{G}(t)=\hat{x}-\hat{k}(t)$
and the reduced Hamiltonians are
\be \label{3.12}
h_s(q,p;t)=\dot{k}(t)\sqrt{m^2+p_1^2+e},\;
\hat{h}_s(\hat{q},\hat{p};t)=\dot{\hat{k}}(t)\sqrt{\hat{p}_1^2-m^2-\hat{e}}
\ee

We now consider the two canonical transformations between these reductions introduced 
above. For the identity map $S=$id we have 
\be \label{3.13}
(\hat{q}^a,\hat{p}_a)=(q^a,p_a)
\ee
so that 
\be \label{3.14}
\hat{h}_s(S(q,p);t)=\dot{\hat{k}}(t)\sqrt{p_1^2-m^2-e}\not=h_s(q,p;T(t))
\ee
Comparing with the first line in (\ref{3.12}) we see that for no reparametrisation $T$ of $t$ 
the two Hamiltonians can be brought into agreement. For the relative
map we compute the relational observables at fixed $t=t_0$ and $\hat{t}=\hat{t}_0$
\ba \label{3.15}
&& Q^a=O_{q^a}(t_0)=q^a+(k(t_0)-x)\frac{p_a}{\sqrt{m^2+p_1^2+E}},\;P_a=O_{p_a}(t_0)=p_a;\; a=1,..,D
\nonumber\\
&& \hat{Q}^1=\hat{O}_{\hat{q}^1}(\hat{t}_0)=\hat{q}^1+(\hat{k}(\hat{t}_0)-\hat{x})\frac{\hat{p}_1}{\sqrt{\hat{p}_1^2 -m^2-\hat{E}}},\;
\nonumber\\
&& \hat{Q}^a=\hat{O}_{\hat{q}^a}(\hat{t}_0)=\hat{q}^a-(\hat{k}(\hat{t}_0)-\hat{x})\frac{\hat{p}_a}{\sqrt{\hat{p}_1^2 -m^2-\hat{E}}};a=2,..,D,\;
\nonumber\\
&& \hat{P}_a=\hat{O}_{\hat{p}_a}(\hat{t}_0)=\hat{p}_a;\; a=1,..,D
\ea
where $E,\hat{E}$ are $e,\hat{e}$ with $p_a\to P_a,\;\hat{p}_a\to \hat{P}_a$. 
Applying $O_\cdot(t_0)$ to the second set of relational observables we must use 
\ba \label{3.16}
\hat{x}=q^1,\;\hat{q}^1=x,\;\hat{q}^a=q^a;\;a=2,..,D;\;
\hat{y}=p_1,\;\hat{p}_1=y,\;\hat{p}_a=p_a;\;a=2,..,D
\ea
Thus at the level of relational observables the map $S$ (we suppress the dependence on $t_0,\hat{t}_0$) is given by 
\ba\label{3.17}  
\hat{Q}^1 &=& k(t_0)+(\hat{k}(\hat{t}_0)-Q^1)\frac{y}{\sqrt{y^2 -m^2-E}}=k(t_0)+(\hat{k}(\hat{t}_0)-Q^1)\frac{\sqrt{m^2+P_1^2+E}}{P_1}
\nonumber\\
\hat{Q}^a &=& Q^a-(\hat{k}(\hat{t}_0)-Q^1)\frac{P_a}{\sqrt{y^2 -m^2-E}}=Q^a+(\hat{k}(\hat{t}_0)-Q^1)\frac{P_a}{P_1};\;a=2,..,D,\;
\nonumber\\
\hat{P}_1 &=& y=-\sqrt{m^2+P_1^2+E},\; \hat{P}_a=P_a;\;a=2,..,D,\
\ea
At the level of true degrees of freedom we just need to replace capital by lower case letters.
Thus
\be \label{3.18}  
\hat{h}_s(S(q,p);t)=\dot{\hat{k}}(t)\sqrt{y^2-m^2-e}=-\dot{\hat{k}}(t)p_1
\ee
Again comparing with the first line in (\ref{3.12}) we see that for no reparametrisation $T$ of $t$ 
the two Hamiltonians can be brought into agreement. In fact it is easy to see that for no 
canonical transformation this will be possible because the square root in $h_s$ is bounded from below
by $m$ while the  square root in $\hat{h}_s$ is bounded from below
by $0$ and no pull back map can change the range of a function.

\subsection{Energy constrained mechanics and the Kepler 2-body problem}
\label{s3.2}  
 
Consider a conserved mechanical system with Hamiltonian $H_0=\frac{1}{2} g^{AB}(K) M_A M_B+U(K)$ for some 
positive definite metric $g$ possibly depending on $K$ and some potential $U$. We fix 
its energy $E$ once and for all. This means we are only considering the solutions of the Hamiltonian equations 
of motion of that energy. We cast this system into the language of section \ref{s2} by introducing the 
totally constrained Hamiltonian $H=V\;Y+X(H_0-E)$ with Lagrange multiplier $V$. This enforces primary constraints 
$Z:=Y=0$ and secondary constraints $C=H_0-E$. This Hamiltonian arises as the incomplete Legendre transform of the Lagrangian
\be \label{3.19}
L(K,U,X,V)=\frac{1}{2X} g_{AB}(K) U^A M^B-X [U(K)-E]
\ee
where $g_{AC} g^{CB}=\delta_A^B$. Note that $E\ge U$ on the constraint surface. Indeed
$M_A=L_{,U^A}=g_{AB} U^B/X,\;Y=L_{,V}=0$ reproduces $H=VY+M_A U^A-L$. Eliminating $X$ from 
$L$ via its Euler Lagrange equation yields the reparametrisation invariant form 
\be \label{3.20}
L(K,U)=\sqrt{\frac{g_{AB} U^A U^B}{2(E-U)}}\
\ee 
for the positive root of $X$. Thus as for the relativistic particle, Lagrangian reparametrisation invariance 
is the gauge symmetry responsible for the secondary constraint and the time variable $t$ becomes a gauge 
parameter. This procedure provides an infinite playground of models in which various technical features 
can be tested.   

As a concrete application consider the 2-body problem in celestial mechanics and focus on the 
relative motion in terms of distance $r$ between point masses and angle $\phi$ in the plane of motion. The Hamiltonian describing 
this system is $H_0=\frac{m}{2}[\dot{r}^2+r^2\;\dot{\phi}^2]+U(r)$ where $m,U$ are effective mass and potential. Equivalently 
$H_0=\frac{1}{2m}[p^2+\frac{l^2}{r^2}]+U(r)$ in terms of momenta $(M_1:=p,M_2:=l)$ conjugate to $K^1:=r,K^2:=\phi$ or 
$H_0=\frac{1}{2m}[M_1^2+M_2^2]+U(r)$ with $r^2=[K^1]^2+[K^2]^2$ in Cartesian coordinates. 
For general energy $E$, which is a constant of motion, we can solve the equations of motion for this
system in three different ways: 0. Use Newton time $T_0=t$ to obtain the trajectory 
$r=r_0(t),\phi=\phi_0(t)$. 
1. Use angular
time $x=\phi$ to obtain the radial shape $r=R(\phi)$. 2. Use radial time $\hat{x}=r$ to obtain the angular shape $\phi=\Phi(r)$. The three formulations are 
of course trivially related by $dr/d\phi=\dot{r}/\dot{\phi}$ where $d/dt(.)=\dot{(.)}$. 

Now fix the energy $E>=0$ and consider 
the constrained system $C:=H_0-E$ describing parabolic or hyperbolic encounters of given energy $E$. The condition $E>=0$ is imposed 
in order that all three time variables are univalent parameters along the entire trajectory. We will work with $r,p,\phi,l$ as 
$K^1, M_1, K^2, M_2$. Following the general reduction programme, we decide on the ingoing $p< 0$ and clockwise $l<0$ branch of motion. 
Then we can equivalently solve the secondary constraint 
$C=0$ using $\bar{C}=l+h(r,p),\;h(r,p)=\sqrt{2m(E-H)r^2+l^2}$ or 
$\hat{\bar{C}}=p+\hat{h}(r,l),\;\hat{h}(r,l)=\sqrt{2m(E-H)+p^2}$. In the notation of the previous section
we have $y=l,x=\phi, q=r,p=p, \hat{y}=p,\hat{x}=r,\hat{q}=\phi,\hat{p}=l$.

We impose gauge conditions 
$G(t):=\phi-k(t),\;\hat{G}(t):=r-\hat{k}(\hat{t})$ respectively and find the 
reduced Hamiltonians 
\be \label{3.21}
h_s(r,p;t)=\dot{k}(t)\;h(r,p),\;\hat{h}_s(\phi,l;\hat{t})=\dot{\hat{k}}(\hat{t})\;\hat{h}(r=\hat{k}(\hat{t}),l)
\ee
This already displays a striking difference between the two reductions: Consider for 
simplicity linear functions $k(t)=ct,\;\hat{k}(t)=\hat{c}t$ for some constants $c,\hat{c}$. 
Then the reduced Hamiltonian of the reduction with respect to to angular time is conservative 
while in the radial time reduction it is explicitly time dependent. 

Again we consider the two canonical transformations of the reduced phase spaces introduced
in the previous section. For the identity reference frame transformation we simply have 
$(\phi,l)=S(r,p)=(r/\ell,p\ell)$ and pulling back the reduced Hamiltonian of the second reduction we find 
\be \label{3.22}  
\hat{h}_s(S(r,p);t)=\dot{k}(t)\;\hat{h}(r=\hat{k}(t),p\ell)
\ee
which obviously is completely different from $h_s(r,p;t)$, no matter which reparametrisation
one might choose, as it does not depend on $r$ at all
but rather retains its explicit time dependence. Note that in this case the identity map is 
at most locally defined because $r/\ell,\phi$ have completely different range. Also we had to introduce 
a length scale $\ell$ in order to have dimensional match without destroying the canonical transformation 
character.  

For the relational reference frame 
transformation we must compute the relational observables. Rather than working 
out the observable maps directly using iterated Poisson brackets it turns out to be shorter 
to proceed as follows: We begin with the observable map $\hat{O}_\cdot(\hat{t})$ and are interested 
first in $\hat{\Phi}:=\hat{O}_\phi(\hat{t})$. It is a Taylor expansion in $-\hat{G}(\hat{t})=\hat{k}(\hat{t})-r$ with 
zeroth order term $\phi$. We note that all iterated Poisson brackets between $\phi$  and $\hat{\bar{C}}$ just display a dependence
on $r,l$. Therefore $hat{\Phi}=\phi+f(r,l)$ for some $f$. We now just need to solve $\{\hat{\bar{C}},\hat{Phi}\}=0$ 
i.e. $-\frac{l}{r^2\;\sqrt{2m(E-U)-l^2/r^2}}+f'(r,l)=0$ where $(.)'=\partial_r(.)$. 
This is because the Taylor series is uniquely determined by this condition and that it equals 
$\phi$ when $r=\hat{k}(\hat{t})$. We define 
$2m U/l^2=:-2/(r_0(l) r), \;2mE/l^2=:1/r_1(l)^2-1/r_0(l)^2$ with $0<r_1(l)<r_0(l)$ depending 
only on $l$ ($m,E$ and Newton's constant are considered fixed) and remember that $l<0$.   
Then $[\frac{1}{\sqrt{r_1^{-2}-(r^{-1}-r_0^{-1})^2}}+f]'=0$ which is solved by 
$f=f_0-\arccos(r_1(1/r-1/r_0))$ where $f_0$ is an integration constant. It is determined 
by the condition that $\hat{\Phi}=\phi$ when $r=\hat{k}(\hat{t}_0)$. Thus 
\be \label{3.23}
\hat{\Phi}:=\hat{O}_\phi(\hat{t})=\phi-\arccos(r_1(1/r-1/r_0))+\arccos(r_1(1/\hat{k}(\hat{t})-1/r_0))
\ee
Taylor expansion of the arcos function at $r=\hat{k}(\hat{t})$ yields the iterated Poisson brackets 
that define the observable map. Since $l$ is already a Dirac observable we trivially have 
\be \label{3.24}
\hat{L}:=\hat{O}_l(\hat{t})=l
\ee
Note that $\Phi$ depends on $r,\phi$ and $l$ (through $r_0,r_1$) but not on $p$. However,
on the constraint surface $l=-h(r,p)$. By construction
\be \label{3.25}
\hat{R}\:=\hat{O}_r(\hat{t})=\hat{k}(\hat{t}), \;\hat{P}:=\hat{O}_p(\hat{t})=-\hat{O}_{\hat{h}}(\hat{t})=
-\hat{h}(\hat{R},\hat{L})
\ee
 
In order to work out $O_\cdot(t)$ we use that according to (\ref{3.23}) the combination 
$g:=\phi-k(t)-\arccos(r_1/(1/r-1/r_0))$ is a Dirac observable because $r_1(l),r_0(l)$ are 
Dirac observables and $k(t)$ is a numerical constant. Thus with $F=r_1(1/r-1/r_0)$
\be \label{3.26}
\cos(g)=F+F(\cos(\phi-k(t))-1)+\sqrt{1-F^2}\sin(\phi-k(t))
\ee 
is a Dirac observable. Since $r_1,r_0$ depend only on the Dirac observable $l$, also 
\be \label{3.27}
\frac{r_1}{r_1/r_0+\cos(g)}=
\frac{r_1}{r_1/r+F(\cos(\phi-k(t))-1)+\sqrt{1-F^2}\sin(\phi-k(t))}
\ee
is a Dirac observable. Note that $b:=r_0/(1+r_1/r_0)$ is the impact parameter or perihel 
and $e=r_0/r_1$ is related to eccentricity, thus we recover the Kepler shape. Both depend on $l$ only. 
Now the right hand side of (\ref{3.26}) equals $r$ for 
$\phi=k(t)$. It follows that 
\be \label{3.28}
R:=O_r(t)=\frac{r_1}{r_1/r+F(\cos(\phi-k(t))-1)+\sqrt{1-F^2}\sin(\phi-k(t))}
\ee
which depends on $\phi,r,l$ only. The Taylor series in terms of $-\phi+k(t)$ gives directly the 
iterated Poisson brackets. We of course have again that  
\be \label{3.29}
L:=O_l(t)=l
\ee
and by construction 
\be \label{3.30}
\Phi:=O_\phi(t)=k(t),\; P:=O_p(t)=-O_{\hat{h}}(t)=-\hat{h}(R,L)
\ee

In the first reduction we use the conjugate pair $Q=R,P=P$ as basic relational observables 
and in the second $\hat{Q}=\hat{\Phi},\hat{P}=\hat{L}$. Then the relational reference frame 
transformation $(\hat{\Phi},\hat{L})=S_{t_0,\hat{t}_0}(R,P)$ at fixed $t=t_0,\hat{t}=\hat{t}_0$ is 
given by 
\be \label{3.31}
\hat{\Phi}=O_{\hat{\Phi}}(t_0)=k(t_0)-\arccos(r_1(1/R-1/r_0))+\arccos(r_1(1/\hat{k}(\hat{t_0})-1/r_0)),\;
\hat{L}=O_{\hat{L}}(t_0)=L=-h(R,P)
\ee
where we use the second relation to write $r_0(l)=r_0(L)=r_0(-h(R,P)):=R_0(R,P)$ and 
$r_1(l)=r_1(L)=r_1(-h(R,P))=:R_1(R,P)$. Here $R=O_r(t_0), P=O_p(t_0)$.

Using the canonical identification 
$(R,P)\to (r,p)$ at fixed $t=t_0$ and  
$(\hat{\Phi},\hat{L})\to (\phi,l)$ 
at fixed $\hat{t}=\hat{t_0}$ of the two independent 
relational observables with the two degrees of freedom of the respective reduction 
we obtain 
\be \label{3.32}
(\phi,l)=S_{t_0,\hat{t}_0}(r,p)=(k(t_0)-\arccos(r_1(1/r-1/r_0))+\arccos(r_1(1/\hat{k}(\hat{t}_0)-1/r_0)),\;-h(r,p))
\ee 
where we substitute $l=-h(r,p)$ for $l$ in $r_0(l),r_1(l)$. Pulling back $\hat{h}_s(\phi,l;\hat{t})$ by 
$S_{t_0,\hat{t}_0}$ we find 
\be \label{3.33}
\hat{h}_s(S_{t_0,\hat{t}_0}(r,p);\hat{t})=\dot{\hat{k}}(t)\hat{h}(\hat{k}(t),-h(r,p))
\ee
Inserting the explicit expressions we find 
\ba \label{3.34}
\hat{h}_s(S_{t_0,\hat{t}_0}(r,p);\hat{t}) &=&
\dot{\hat{k}}(\hat{t})\sqrt{2m(E-U)(\hat{k}(\hat{t}))-[2m(E-U(r))-p^2]r^2/\hat{k}(\hat{t})^2},\;
\nonumber\\
h_s(r,p;t) &=& \dot{k}(t)\sqrt{[2m(E-U(r))-p^2] r^2}
\ea
which are also very different from each other. Pick again $k(t)=ct,\hat{k}(t)=c't$ and $t_0=\hat{t}_0=0$ then 
the first retains its explicit time dependence while the second is conservative. Moreover,
$p^2$ enters the two square roots with opposite signs. No reparametrisation can change this.   
Finally note that the square roots have different dimension, namely linear and angular 
momentum respectively. This can be adjusted by choosing the dimensions of $c,\hat{c}=c/\ell$ to be 
that of velocity and frequency respectively.

\subsection{Discussion}
\label{s3.3}

These counter examples reveal that under the two natural canonical transformations of the 
reduced phase spaces that we identified, the two reduced Hamiltonians are not images of each other
for no choice of reparametrisation. The 
fact that the reduced phase spaces in terms of true degrees of freedom are Poisson isomorphic 
with relational observables and physical Hamiltonians means that this is by no means 
an indication of loss of gauge invariance. There can be no doubt that the description 
of the reduced phase space and its dynamics in terms of relational observables and physical Hamiltonians  
is a valid description of gauge invariant content of the system, irrespective of which 
choice of reference fields one uses in order to define them. For instance, both reductions in the Kepler 
problem yield correctly the Kepler parabola and hyperbola, no matter whether one uses angle or radius 
as reference field, however, one must use effective or reduced Hamiltonians which are not naturally 
images of each other under these canonical transformations. Considering the canonical transformations 
introduced, we consider the relational reference frame transformation as the 
physically distinguished one as it respects the ranges and dimensions of canonical coordinates and 
simply accounts for the fact that two sets of relational observables are algebraically dependent 
on each other in a very definite way. We saw already in these basic examples that these algebraic relations 
are extremely non-linear due to the square roots that necessarily enter the transformations in realistic 
models (i.e. those which correctly capture the fact that the Hamiltonian constraint is a quadratic 
expression in the momenta).

That the relational reference frame transformation, which is explicitly time dependent, changes 
the physical Hamiltonian is perhaps intuitively not surprising because in general explicitly time dependent 
canonical transformations change the Hamiltonian by a term that depends on the generator of 
that transformation. That the identity reference frame transformation changes the Hamiltonian,
which is not explicitly time dependent, cannot be understood in this way. Moreover, as we 
will describe in more detail in section \ref{s8}, in situations where the physical Hamiltonian 
can be chosen to be a boundary term, even the relational reference transformation is not 
explicitly time dependent and hence the change of Hamiltonians can also not be understood in this 
way.  

The technical reason why the physical Hamiltonians with respect to different reference frames 
are not related by pullback under the relational reference frame transformation map despite it being 
a canonical transformation is as follows: The physical Hamiltonian in a given reference frame 
defined by the reference field $x$ arises from the observable map. It consists in 1. performing the canonical transformation generated 
by $g\bar{C},\;\bar{C}=y+h(x;q,p)$ and 2. equating at $g=k(t)-x$. This is not the same 
as 3. performing the canonical transformation generated by $\bar{C}(t):=(k(t)-x)\bar{C}$ which is
a strange gauge transformation since the gauge parameter is phase space dependent.  
The transformation 3. is canonical on functions of the full kinematical phase space 
with generator $(k(t)-x)\bar{C}$ which acts trivially on Dirac observables 
while the combination of 1. and 2. is only canonical 
on functions of the true degrees of freedom $q,p$ (because it projects non gauge invariant functions to 
gauge invariant ones thus cannot have full range) 
and has as generator the physical Hamiltonian 
$H_s(Q,P,t)=\dot{k}(t) h(k(t),Q,P)$ where $Q=O_q(0),P=O_p(0)$ which acts non-trivially on Dirac observables.  

In fact, the reduced or physical Hamiltonian must depend on the reference frame because 
it is the effective Hamiltonian which generates the same dynamics for the true degrees of 
freedom as the primary Hamiltonian does, when constraints, gauge fixing conditions and 
stability conditions are satisfied. Hence it preserves the reference frame, in particular 
the gauge fixing conditions on the reference field, and thus depends 
non-trivially on it.  

The dependence of a Hamiltonian on a reference frame is not unfamiliar. Even in free 
field theory on Minkowski space, the Hamiltonian does depend on the inertial frame. This
is because the Hamiltonian is no Lorentz scalar but a component of the energy momentum 4-vector. 
Thus under change of inertial frame, it changes by the corresponding Lorentz transformation.
The purpose of that transformation in that case is to preserve the foliation of spacetime into leaves of 
constant time, and what one calls time depends on the reference frame. Of course 
the Lorentz transformation between the frames is implementable as a canonical transformation
just as in the general case the relational reference frame transformation is, which in this case even can be promoted to a unitary 
transformation in the quantum theory. However the Hamiltonian to use in the boosted frame 
is not just the pull-back of the Hamiltonian in the original frame, rather it is a linear 
combination of Hamiltonian and momentum in boost direction of that frame.  
We will rederive this well known fact from the perspective 
of generally covariant gauge systems in the appendix. In that particular case, the failure 
of the physical Hamiltonian of the second reference frame from reproducing the 
physical Hamiltonian of the first reference frame upon pullback by the relational reference frame transformation,
which of course is always a Dirac observable, has an immediate physical interpretation in terms of 
the momentum in boost direction of the first reference frame. 

The only possibility that the Hamiltonians in the two frames do coincide, i.e. are 
pull-backs of each other under the relational reference frame transformation and a reparametrisation, is that 
the secondary constraint $C$ itself has a very special type of symmetry with regard to the split into 
true and gauge degrees of freedom. Working again with $q,p,x,y$, consider some function $E=E(q,p)$ 
and the constraint $C(q,p,x,y)=E(q,p)-E(x,y)$. Note that the 
dependence of $C$ on $q,p$ and $x,y$ is via the same function $E$. Therefore we can also solve 
$C=0$ for $y$ or $p$ using the same function $h$ i.e. $\bar{C}=y+h(x;q,p)=0$ and $\hat{\bar{C}}=p+\hat{h}(q;x,y)$ 
with $h=\hat{h}$ and the 
reduced Hamiltonians are $H=\dot{k}(t)\;h(k(t),q,p)$ and $\hat{H}=\dot{\hat{k}}(\hat{t})\hat{h}(\hat{k}(\hat{t}),x,y)$. 
In this case the identity canonical transformation $(x,y)=S(q,p):=(q,p)$ and the reparametrisation $\hat{t}=T(t)$ such that 
$k=\hat{k}\circ T$ yields due to $h=\hat{h}$ that we have $dt H=d\hat{t} \hat{H}_{\hat{t}=T(t),(x,y)=S(q,p)}$.
However note that even in this case the identity map is not granted to be the relative reference frame exchange map.
A trivial example would be $E(q,p)=p^2/2$. Then $h(x,;q,p)=p$ and we have the branch constraint $\bar{C}=\hat{\bar{C}}=y+p$
and relational observables $O_q=q+k-x, \; O_p=p,\; \hat{O}_x=x+\hat{k}-q,\; \hat{O}_y=y$. Thus 
$S(O_q,O_p):=(O_{\hat{O}_x},O_{\hat{O}_y})=(k+\hat{k}-O_q,-O_p)$. Hence in terms of the true degrees of 
freedom $S(q,p)=(k+\hat{k}-q,-p)$ is a canonical transformation but not the identity transformation. Still 
in this case the reduced Hamiltonians
$H=\dot{k}p,\;\dot{\hat{k}}y$ can be brought to match under pullback by $S$ and the reparametrisation
$\hat{t}=T(t)$ such that $\hat{k}\circ T=-k$. It transpires that cases with matching Hamiltonians 
after canonical transformations with respect to the reference frame exchange map and reparametrisations 
are rather exceptional and artificial. Note that the identity of Hamiltonians after pull-back is 
achieved despeite the fact that the transformation is explicitly time dependent, hence the fact 
that the transformation is explicitly time dependent does not imply the change of the Hamiltonian under pull-back
either.

\section{Changing reference frames, the fluctuation paradox and quantum clocks}
\label{s4}      

The fluctuation paradox refers to the fact that if one uses a certain field $x$ as a reference field 
in one reference frame but another field $\hat{x}$ as a clock in another reference frame, then 
in the first frame the relational observable associated to $x$  becomes proportional 
to the unity function $1$ on the phase space while in the second the relational observable 
associated to $x$ retains a non-trivial phase space dependence. Therefore upon quantisation 
the first relational observable has zero fluctuations in any representation of the 
canonical commutation and adjointness relations as it is quantised as 
proportional to the identity operator while in the second it has non-trivial fluctuations.
As the quantum fluctuations of observables, i.e. gauge invariant objects which relational 
observables certainly are, can certainly measured, there is apparently 
an obvious problem as it indicates a loss equivalence of reference frames. This problem cannot 
be avoided by mapping by unitary operators, which are the quantum analogs of classical 
canonical transformations which always map the unit operator to itself. 

The resolution of this paradox is in fact quite simple and follows from the 
corresponding relational reference frame transformation 
which is central to this work. We follow the general framework developed and 
consider the secondary constraint $C=0$ equivalently be solved by $\bar{C}=y+h(x;q,p)$ or 
$\hat{\bar{C}}=\hat{y}+\hat{h}(\hat{x};\hat{q},\hat{p})$ respectively with respect to two
different splits $(K,M)=((x,y),(q,p))=((\hat{x},\hat{y}),(\hat{q},\hat{p}))$ of the 
secondary phase space with mixtures among the gauge and true degrees of freedom. To make it 
concrete, consider the case of a single constraint and the situation $x=\hat{q},y=\hat{p},
q=\hat{x},p=\hat{y}$. Let 
\be \label{4.1}
f\mapsto O_f(t)=[e^{g\{\bar{C},.\}}\;\cdot f]_{g=k(t)-x},\;\;
f\mapsto \hat{O}_f(\hat{t})=[e^{g\{\hat{\bar{C}},.\}}\;\cdot f]_{g=\hat{k}(\hat{t})-\hat{x}},\;\;
\ee
be the corresponding observable maps. Then in our now familiar notation
\be \label{4.2}
X(t):=O_x(t)=k(t)\;1, \;\;\hat{X}(\hat{t}):=\hat{O}_x(\hat{t})=\hat{O}_{\hat{q}}(\hat{t})=:\hat{Q}(\hat{t})\not\propto 1
\ee
This is a mathematical fact. However, the point is now that there is no contradiction:
The trivial relational observable $X(t)$ is {\it physically different} from 
the non trivial relational observable $\hat{X}(t)$. The interpretation of $X(t)$ is 
that Dirac observable which has the property that it takes the value 
of $x$ in the gauge when $x$ takes the value $k(t)$ which not surprisingly yields 
$k(t)$. The interpretation of $\hat{X}(t)$ is that it is that Dirac observable which has the property 
that it takes the value of $x$ when $\hat{x}$ takes 
the value $\hat{k}(\hat{t})$ which is an entirely different definition and thus yields a different 
Dirac observable.

As already stated many times throughout this paper, the dimension of the reduced phase space 
is independent of the choice of reference frame. For every choice of reference frame and every choice of time, there 
exists a set of elementary relational Dirac observables which all others are 
functions of. Using the fact that relational observables corresponding to the same 
reference frame at different times are uniquely related by the dynamics driven by the 
the corresponding physical Hamiltonian and initial data, we can focus on the case 
of tow fixed times $t_0,\hat{t}_0$. For the reference frame $(x,y)$ these elementary relational observables 
can be chosen as the relational observables 
$Q:=O_q(t_0),\;P=O_p(t_0)$ and for the reference frame $(\hat{x},\hat{y})$ these can be chosen as  
the relational observables $\hat{Q}=\hat{O}_{\hat{q}}(\hat{t}_0),\hat{P}=\hat{O}_{\hat{p}}(\hat{t}_0)$
where $t_0,\hat{t}_0$ are any two fixed values (e.g. zero). Therefore, the functions 
$\hat{Q},\hat{P}$ must be concrete functions of the $Q,P$. This functional relation 
is precisely captured by the relational reference frame transformation
\be \label{4.3}
(\hat{Q},\hat{P})=S(Q,P):=(O_{\hat{Q}}(t_0),O_{\hat{P}}(t_0))  
\ee
%which is a mathematical identity on the constraint surface. 
This map has been worked out concretely in the above examples, see (\ref{3.17}), (\ref{3.31}).
Coming back to 
the apparent paradox we saw that $\hat{X}=\hat{Q}$ is a non-trivial Dirac 
observable in the second reference frame. That {\it same} Dirac observable 
remains non-trivial when expressed via (\ref{4.3}) in terms of the Dirac observables $Q,P$ of the 
first reference frame. Hence, upon quantisation, in both reference frames these would 
display non-trivial fluctuations. Likewise, the trivial observable $1$ in the first reference 
frame is mapped to $\hat{O}_1(\hat{t}_0)=1$ in the second and thus display zero fluctuations 
upon quantisation in both reference frames.\\
\\
An entirely different subject concerns quantum clocks. By this we mean the quantisations of gauge invariant 
functions that are sensitive to an operationally defined time measurement, say an actual atomic 
clock. This will be a non-trivial operator in any reference frame, built from the relational observables corresponding to standard 
model fields to define the Rabi oscillations of a Caesium atom exposed to an electromagnetic 
field. This concept fits within the branch of quantum physics in which the measurement apparatus is 
part of the quantum system rather than a classical external observer and is not in conflict 
with the relational observable corresponding to the reference field that defines the reference frame 
as being quantised by the unit operator. Thus a quantum clock, from the point of view of 
relational reference frames is to be understood in the usual way, see e.g. \cite{8b}. 

Finally, some authors consider the quantisation of a reference field, rather than its associated relational observable,
on the kinematical Hilbert space as a quantum clock which certainly does have fluctuations. 
But based on the premise that only gauge invariant functions or operators correspond to classical
or quantum observables, such a definition of a quantum clock is physically meaningless because 
the reference field itself, rather than the corresponding relational observable, is not gauge invariant,
in particular its fluctuations have no physical interpretation. The only sensible role that such a quantum
reference field can play is in the quantisation before reduction approach where it enters the quantum 
constraint operator and the relational observables considered as operators on the kinematical Hilbert space
(an restricted to the generalised joint kernel of the constraints).
But again that quantum reference field is not observable. Only the compound operator built from 
quantum reference fields and quantum true degrees of freedom entering a relational observable can be
gauge invariant and thus observable. Then a remaining meaningful question to ask is whether quantisation before 
and after reduction are equivalent with respect to those observables. This is rather unlikely for a general 
system, there is no unique road to quantisation of a classical system and poses again no conceptual problem.

\section{Interpretation, measurement and relation to observation}
\label{s6}

We have established that the description of the system in terms of true degrees of freedom 
defined by a certain choice of reference frame is mathematically equivalent to the description 
in terms of the relational observables defined by the same reference frame, both kinematically and 
dynamically. Now a basic postulate in gauge theories is that experiments have only access to gauge 
invariant functions on the phase space that are called observables. Thus, making use of the mathematical 
equivalence between true degrees of freedom and relational observables, the true degrees of freedom are 
measurable when interpreted as relational observables that are gauge invariant by construction. Moreover,
while one may use different reference frames in order to define relational observables, two sets of relational 
observables obtained from two different reference frames are functions of each other as we have seen.
Therefore using the true degrees of freedom description is just for mathematical convenience, the 
proper interpretation is in terms of the corresponding relational observables.

Having settled this, we now come back to the list of questions of the introduction. Taking the example 
of General Relativity, the central object of interest is the metric tensor field. Since we can certainly 
measure all of its ten components, by the basic postulate all components must be observables. 
However, a metric tensor field is subject to gauge transformations due to diffeomorphism invariance.
The apparent contradiction can only be removed by interpreting the ten components of the metric 
tensor as expressed as functions of the relational observables defined by a reference frame.

Explicitly, this works as follows: In our notation, the fields $x^I, q^a$ stand for fields
that encode the spatial-spatial components of the metric field while the fields $X^I$ encode the 
temporal-temporal and temporal-spatial components (lapse and shift) of the metric field. Now recall that 
picking $x^I$ as reference fields, they get fixed to take numerical values, i.e. constants 
with respect to their dependence on the true degrees of freedom, the $q^a$ become
true degrees of freedom and the $X^I$ become also functions of the true degrees of freedom 
$q^a, p_a$. Thus indeed all ten components of the metric tensor field have been written 
as functions of the true degrees of freedom $q^a,p_a$ which encode the two independent polarisations 
of the metric tensor field. The dynamics of all ten components then follows from dynamics of the 
true degrees of freedom with respect to the reduced Hamiltonian (and the explicit time dependence 
encoded in $k^I(t)$). When reinterpreted as relational Dirac observables, we thus have resolved the puzzle how 
gauge invariance, observability and measurability of a priori gauge dependent fields can be reconciled.

Now suppose that a certain reference frame and gauge was fixed. E.g. one could decompose the three metric into 
a conformal factor and a uni-modular metric and impose that the conformal factor be only a function of 
of time (one gauge condition per spatial point) and that the unimodular part be transverse with respect 
to some time independent background metric (three gauge conditions per spatial point). Then an experiment 
measuring the metric may come to results that contradict this gauge condition. The interpretation must be 
that since the measurement must have been in {\it some} reference frame, the theoretical calculation 
is performed in a reference frame that the laboratory in which the measurement was performed is not. 
Then one has two options, adapting the theory frame or adapting the laboratory frame.
From the theory side, either one performs sufficient measurements in order to find out what reference 
frame the laboratory frame is in and then recalculates everything using this experimentally determined 
reference frame to formulate the corresponding gauge condition or one applies the relational reference frame transformation
between theory frame and measured frame to one's calculation in order to be able to compare the data.
From the experimental side, one uses the spacetime diffeomorphism that relates the reference frames 
in order to express the laboratory data in the theory frame. 

Finally we note that since the relational observables defined by two different reference frames are 
functions of each other, one might as well use the physical Hamiltonian of the second reference frame to  
evolve the relational observables defined by the first frame. This is true but it would 
lead to the wrong dynamical law. This is because,  
as we have seen in section \ref{s2}, it is only the reduced Hamiltonian determined by a given 
reference frame that has the property to preserve that frame (stability of the gauge condition). Therefore a set 
of relational observables and its physical Hamiltonian are tied to each other by the choice of reference 
frame because by design the dynamics must respect the gauge condition or reference frame.

\section{Quantum relational reference frames and quantum relational reference frame transformations}
\label{s7}

The considerations so far were entirely classical but of course they prepare for 
reduced phase space quantisation. The first thing to worry about is how to 
deal with the concrete implementation in terms of branches of the constraint surface.
See \cite{8z} and references therein for a model with finitely many degrees of freedom where 
a similar branch analysis in the quantisation before reduction approach was carried out.
With respect to a given choice of reference fields $x^I,y_I$ we solved the constraints 
$C_I$ in the form $y_I+h_I(x;q,p)=0$ where some roots $h_I$ come with 
a sign ambiguity while $q^a,p_a$ typically have full range in a co-tangent bundle. Let us 
encode that sign ambiguity as $h_I^{\sigma_I},\; \sigma_I=\pm$. We allow the case that 
there is in fact only one root for a particular index $I$ by simply setting $h_I^+=h_I^-=h_I$
in that case. The case of a single root in fact appears in the cases of interest arising from the 
constraints that are linear in the $y_I$ to begin with.
In general the roots are not simply related by $h_I^\sigma=\sigma h_I$ for $h_I$ independent
of $\sigma$. Thus there are up to $2^N$ possible branches where $N$ is the number 
of distinct roots and typically $N$ is infinite in field 
theory. We can display a branch as the sequence $\sigma=(\sigma_I)_I$.  

On each branch $\sigma$ one could in principle pick a different gauge fixing condition
$x^I=k^I_\sigma(t)$. If the number of possible branches is infinite one can use physical arguments in order to select one or at least 
a finite number of those branches as distinguished, such as positivity of energy 
$h(t;q,p)=\sum_I \dot{k}^I_\sigma(t) h_I^{\sigma_I}(k_sigma(t),q,p)>0$ arguments or continuity arguments
across branches. Another possibility is to use the fact 
that the entire constraint surface defined by $C_I$ is a disjoint (up to lower 
dimensional boundaries)
union of branches. If $C_I$ is 
is quadratic in the $y_J$ then $C_I$ is of the form 
$C_I=M_I^{JK}\; (y_J+h^+_J)\;(y_K+h_K^-)$ for suitable, phase space dependent 
matrix valued functions 
$M_I^{JK}$. Their form decides on which sign configuration $\sigma$ is an allowed branch
$\sigma$: for each allowed $\sigma$ the $M_I ^{JK}$ are such that $C_I=0$ for all
$I$ when $y_I+h^{\sigma_I}_I=0$ for all $I$ and conversely $C_I=0$ for all $I$ 
implies that $y_I+h^{\sigma_I}_I=0$ for all $I$ for one of the allowed $\sigma$. 
The extreme case is that $M_I^{JK}=0$ for $J\not=K$ in which case all $\sigma$ are allowed
and $M_I^K:=M_I\;^{JJ}$ is invertible on the subset of constraints quadratic in the 
$y_K$. The continuous choices $\sigma_I=+$ or $\sigma_I=-$ for all $I$ is always allowed and 
leads to this restricted form of $C_I$ (i.e. terms $\propto (y_J+h^\pm_J)(y_K+h^\pm_K)$ are 
not allowed). We see that the gauge evolution of $x^I$ stops when branches 
cross at $y_I=-h_I^+=-h_I^-$ (if $h_I^+=h_I^-$ is at all possible) for all $I$. 
Thus the branches can be considered as independent systems and the reduced 
phase spaces on each branch are all isomorphic to the reduced phase space defined 
by the true observables $q,p$. If the phase space of the true observables is a
standard cotangent bundle over a configuration space, a possible quantisation of the entire system
consists in a direct sum of Hilbert spaces  
\be \label{7.1}
{\cal H}=\oplus_\sigma {\cal H}_\sigma,\; \sigma=\{\sigma_I\}
\ee
where $({\cal H}_\sigma,\rho_\sigma)$ is a representation of the Weyl algebra $\mathfrak{A}$ 
(paying attention to the topology of the configuration space in which $q$ takes values) 
defined by $q,p$ such that some ordering of 
\be \label{7.2}
H_\sigma(t,q,p):=\sum_I\; \sigma_I\; \dot{k}^I_\sigma(t) \; h_I^{\sigma_I}(k(t),q,p)
\ee
is self-adjoint on ${\cal H}_\sigma$. In the special case that $h_I^{\sigma_I}=\sigma_I \; |h_I|$
one could obtain positivity of (\ref{7.2}) for the choice 
$k_I^\sigma(t)=\sigma_I\; k_I(t),\; \dot{k}_I(t)\ge 0$ and $H_\sigma$ would be branch independent.
In the unlikely case that all relational observables are block diagonal with respect to the direct 
sum structure, the branch sectors labelled by $\sigma$ are super selected. 

Note that the introduction of the branches cannot be avoided unless the kinematical phase space 
admits a polarisation in which the momenta appear only linearly which is not the case for the 
gauge systems of interest. Then it is a fact that the constraint surface has a branched structure
and since the full reduction of the system requires to solve the constraints, we have to face 
this complication. One may consider working with the alternative constraints $\tilde{C}_I=
(\Delta^{-1})_I^J C_J$ as long as the matrix $\Delta_I^J=\{C_I,x^J\}$ is not singular at least 
in a neighbourhood of the constraint surface and then construct the observable map 
$O_f(t)=[\exp(\{g^I \tilde{C}_I,.\})\cdot f]_{g=-G(t)}$ thus effectively replacing 
$\bar{C}_I=y_I+h_I^{\sigma_I}$ used in the main text by $\tilde{C}_I$. Indeed the properties 1.-6. quoted in section
\ref{s2} continue to hold weakly \cite{3c,3d} and $\tilde{C}_I$ is apparently free of the branch discussion.
However, $\Delta$ becomes singular precisely when branches cross. Moreover, in a reduction after 
quantisation approach one employs a rigging map (also called group averaging map, see e.g. \cite{8a}) 
that maps vectors on the kinematical Hilbert space 
to generalised zero eigenvectors of all constraints. This is formally given by $\Psi\mapsto \delta(C)\Psi:=\prod_I \delta(C_I) \Psi$,
assuming that the $C_I$ are self-adjoint on the kinematical Hilbert space so that the $\delta$ distributions
can be defined by the spectral theorem. There are two problems with this definition. First, unless 
the $C_I$ form at least a Lie algebra (with structure constants rather than structure functions - the 
naive product of $\delta$ distributions is then replaced by an invariant integral over the gauge group) this  
procedure does not produce an element in the joint kernel. Since this is not our situation,
we must treat the $C_I$ differently. Second, even when formally computing the physical 
inner product by $<\delta(C)\Psi,\delta(C)\Psi'>_{{\sf phys}}:=<\Psi,\delta(C)\Psi'>_{{\sf kin}}$
at some point one must solve the $\delta$ distributions which reintroduces the branches. Therefore 
it appears that 
\be \label{7.3}
\delta(C):=\sum_\sigma \prod_I\delta(y_I+\sigma_I h^{\sigma_I}_I)
\ee
is the more promising way to define the rigging map thus reintroducing the branches because the 
classical constraints $\bar{C}_I=y_I +\sigma_I h_I$ are Abelian with respect to the Poisson bracket
and thus there is a chance that the constraints can be implemented as commuting operators.
Note that the decomposition of the full $\delta(C)$ distribution into a sum over branches formally leads to a physical inner 
product by the rigging map in agreement with the direct sum structure (\ref{7.1}) that we reached at 
by the reduction before quantisation approach.

Now we consider the change of reference frames $(x,y)\to (\hat{x},\hat{y})$. Several interesting 
further issues now arise. 
First, from the classical perspective, it could happen that the two reference frames 
partition the entire secondary constraint surface $\bar{\Gamma}_s$ into different branches
$\bar{\Gamma}_{s,\sigma},\;\hat{\bar{\Gamma}}_{s,\hat{\sigma}}$ in the sense that for given $\sigma$ we have
$\bar{\Gamma}_{s,\sigma}\cap\hat{\bar{\Gamma}}_{s,\hat{\sigma}}\not=\emptyset$ for more than one $\hat{\sigma}$. 
We have seen in the examples that this is what typically happens (e.g. in the Kepler problem we can have 
four situations: in and outgoing (sign of radial momentum $p$) as well as clockwise and counter clockwise (sign of 
angular momentum) orbits). Then the relational reference frame transformation $S_{t,\hat{t}}$ for given $t,\hat{t}$ 
is a priori only defined 
on those non-empty intersections $\bar{\Gamma}_{s,\sigma}\cap\hat{\bar{\Gamma}}_{s,\hat{\sigma}}\not=\emptyset$ and 
as such is only locally a canonical transformation because some of the $p,y$ respectively are some of the $\hat{y},\hat{q}$ respectively 
hence if the $y,\hat{y}$ have prescribed signs, some of the $p,\hat{p}$ also have prescribed signs so that the reduced 
phase spaces restricted to intersections are not full co-tangent bundles. The relational reference frame 
transformation may or 
or may not be extendible by continuity to all of $\bar{\Gamma}_s$. At least in the examples however, we saw that 
such an extension is indeed possible and it is likely to be possible in general because the branches descend from 
the branch free functions $C_I$. 

Supposing that this is the case, one can ask whether one can implement the relational reference 
frame transformation as a unitary transformation between the corresponding Hilbert spaces and Weyl algebras given that 
their classical analog is a unitary transformation. We saw already in simple examples that this canonical transformation 
is rather non-linear, thus one expects a quantisation of the relational reference frame transformation as a unitary transformation between 
the two direct sums of representations of Weyl algebras to be technically challenging in the generic case. 
In that respect it is interesting to recall the results from \cite{9} that even within the context of free Quantum fields 
on Minkowski spacetime a unitary transformation between the quantum field theories defined for inertial and non-inertial 
foliations may fail to exist. This problem is in principle much simpler because one did not switch reference fields but 
only gauge conditions (see appendix), when one switches both, a much higher degree of non-linearity appears. \\
\\
Supposing that the corresponding quantum relational reference frame transformation is implementable as a unitary transformation
one can ask what it has to do  
with the bulk of the literature on quantum reference frames which so far concerns mostly systems with finitely many 
degrees of freedom and without gauge redundancy. In what follows, we describe concrete possibilities for
such a link.

Much of the work on quantum reference frames studies the following abstract setup: There is a  
tripartite quantum system described by three sets A,B,C of independent observables. 
Note the term "observable" - the system is assumed to be already reduced and all potential gauge redundancies are assumed 
to be taken care of. The observables in the respective set A,B,C are interpreted as intrinsic to a corresponding 
subsystem also labelled A,B,C. Each subsystem has access to all observables. 
Being independent,
the observables of the three sets are represented as self-adjoint operators on their own 
independent Hilbert space ${\cal H}_{{\sf I}},\;{\sf I}=$A,B,C and the total system is 
described by the tensor product ${\cal H}=\otimes_{\sf I}\;{\cal H}_{{\sf I}}$. 
One can now construct
from the intrinsic observables, which operate factor wise on the tensor product, composite ones 
which no longer act factor wise. In particular, one can construct such observables which 
one can coin "common" and "relative". The example introduced in \cite{4b} is helpful 
establish some intuition: consider three particles A,B,C in classical 
non relativistic mechanics with 
corresponding canonical coordinates $(q^j,p_j);\; j=$A,B,C as the intrinsic observables 
and another canonical chart defined by the common centre of mass coordinate $R=\frac{1}{3}\sum_j q^j$ 
and the relative coordinates $x^j=q^k-q^l;\; \epsilon_{jkl}=1$ together with their conjugate momenta.
(note the identity $\sum x^j=0$). The transition between the intrinsic observables to the 
relative ones is assumed to be a canonical transformation (such as in this example). Then 
for each of the three subsystems, one can consider a different set of relative canonical    
coordinates as "natural" to use. In our example, these could be the coordinates $x_{(A)}^1:=x^B, x_{(A)}^2:=x^C$ for system
A and cyclic. Since the relative coordinates are not independent of each other, for instance  
a pure, separated 
vector state with respect to the relative observables that system A uses is a pure, entangled vector state with respect to 
the relative coordinates that system B uses and cyclic when using the Schr\"odinger picture. 
In the example e.g. this is due to the correlation 
$x_{(A)}^1 -x^1_{(B)}-x^2_{(B)}, x_{(A)}^2=x_{(B)}^1$
%=-(x^C + x^A)=
among operators in the Heisenberg picture.
The classical canonical transformation in the examples 
considered so far in the literature are implementable as unitary transformations. 

Moreover, if different subsystems do not have complete access to all observables of the full system 
then in probability amplitudes one should average over the possible measurement outcomes of the 
non-accessible observables (e.g. in scattering experiments, one sums over individual 
probabilities including polarisations to get the probability for all polarisations if the 
experiment is not sensitive to polarisation). Applied to the present situation  
one should average over the group G of transformations that 
relate intrinsic observables that encode reference frame information that another reference frame 
has no knowledge about. This operation has been coined G-twirl in \cite{4c}, see 
\cite{8c} for an extension to general frame groups,
but it is 
identical to the above mentioned group averaging tool employed in the constrained 
Hamiltonian systems, see \cite{8a} and references therein. 
%which unsurpringsingly produces mixed states from pure, superselection and decoherence

We can now establish how these considerations fit into the language of relational reference 
frames and relational reference frame transformations, see also 
\cite{8d} for related ideas for finite dimensional deparametrising systems: First of 
all, the above considerations all take place in terms of the observables of the theory. 
Therefore these can only be implementable in terms of relational observables. 
In case that above considerations concern only a finite number of degrees of freedom,
the corresponding observables will correspond to complicated compound observables built 
from the infinitely many field theory observables.
There are then
two possibilities. Either the observables considered above are the relational observables 
corresponding to the true canonical coordinates of just one relational reference frame 
or of more than one. In the first case, above
considerations are simply extra structure that one can consider after the gauge reduction 
has been performed for one and the same choice of reference field and gauge conditions thereon
and nothing is changed in the above considerations. 

More likely is the second possibility that the observables 
considered above involve relational observables corresponding to canonical coordinates of 
several relational reference frames. Concretely, as we did in section \ref{s3}, one can consider a split of the 
set of all kinematical canonical pairs $(K,M)$ as $((q,p),(x,y))$ where 
$(q,p)=((c,d),(r_1,s_1),(r_2,s_2))$ and $(x,y)=((x_1,y_1),(x_2,y_2))$. Here $q,p$ are the true
degrees of freedom with respect to the reference fields $x,y$ of a given relational reference frame.
Likewise we can consider another split $(K,M)$ as $((\hat{q},\hat{p}),(\hat{x},\hat{y}))$ where 
$(\hat{q},\hat{p})=((c,d),(\hat{r}_1,\hat{s}_1),(\hat{r}_2,\hat{s}_2))$ and 
$(\hat{x},\hat{y})=((\hat{x}_1,\hat{y}_1),(\hat{x}_2,\hat{y}_2))$. We pick these splits such that
$r_1=\hat{r}_1, \;s_1=\hat{s}_1, x_1=\hat{x}_1, \;y_1=\hat{y}_1$ and 
$r_2=\hat{x}_2, \;s_2=\hat{y}_2, x_2=\hat{r}_2, \;y_2=\hat{y}_2$. Let $O,\hat{O}$ be the 
observable maps corresponding to the relational reference frames $(x,k)$ and $(\hat{x},\hat{k})$ and 
$S_{t,\hat{t}}$ the corresponding relational reference frame transformation (say at $t=\hat{t}=0$), i.e. 
$(\hat{O}_q(\hat{t}),\hat{O}_p(\hat{t}))=S_{t,\hat{t}}(O_q(t),O_p(t))$. Suppose that we can arrange 
that this map is 1. block diagonal and 2. its restriction to $O_c(t),O_d(t)$ is the identity map at $t=\hat{t}$.
Then 
$O_c(t),O_d(t)$ can be considered as common degrees of freedom and $O_r(t),O_s(t)$ 
with $(r,s)=((r_1,s_1),(r_2,s_2))$ 
as relative 
degrees of freedom. 
The restriction of $S_{t,\hat{t}}$ to  
$O_r(t),O_s(t)$ is a non-trivial canonical transformation that expresses  
$\hat{O}_{hat{r}}(\hat{t}),O_{\hat{s}}(\hat{t})$ in terms of these 
where $(\hat{r},\hat{s})=((\hat{r}_1,\hat{s}_1),(\hat{r}_2,\hat{s}_2$. Then the relational
reference frame transformation $S_{t,\hat{t}}$ plays the role 
of the canonical transformation between relative coordinates in above considerations. 
We note that in this case the relational reference frame transformation arises due to 
dynamical considerations and is not some extra structure. The  
physical Hamiltonian in both frames is also dictated by underlying dynamics of 
the gauge system and its forms and transformation behaviour need need not be 
derived by independent means.\\
\\
Obviously, all these and related question require a much deeper analysis which is beyond the scope 
of the present manuscript whose primary purpose is to give an overview over the required 
techniques and arising subtleties.

\section{Summary, further observations and outlook}
\label{s8}   

In this work we have addressed several issues that concern the conceptually and technically difficult 
class of theories which are subject to spacetime diffeomorphism gauge invariance. While 
similar individual items of our exposition have surely been discussed elsewhere 
we believe that their particular combination in the present contribution is of some added value in that 
it aims to tie together notions that have appeared in different contexts and communities and 
to arrive at a mathematically and conceptually coherent picture in a concrete and rigid framework 
within which interesting open questions 
can be addressed. The framework is general enough to cover the gauge field theories that 
to date are our best candidates to describe the observations, see \cite{3e} for concrete 
quantum implementations mostly with material reference fields. 
We may summarise this picture, obtained in the 
quantisation after reduction approach, as follows, thereby
taking the opportunity of bundling several 
almost synonymous notions and definitions and embedding them into a wider context: 
\begin{itemize}
\item[0.] {\it Gauge invariance, observables, observations, measurements}\\
A basic assumption, which is therefore mostly not explicitly stated, is that what can be observed or 
measured in nature must not suffer from any arbitrariness. Since in a gauge theory there is 
arbitrariness in that one can perform gauge transformations on the system whose dynamics 
cannot be predicted, one postulates that only gauge invariant objects can be observed.
These gauge invariants are called observables.
\item[1.] {\it Reference fields, gauge conditions, relational reference frames}\\
A reference field is a field which is used in order to construct gauge invariant, hence observable, fields 
from gauge dependent ones. Technically this works by imposing gauge conditions on the reference 
fields. A choice of a pair of reference fields and gauge conditions thereon determine a 
relational reference frame, i.e. equip a priori arbitrary 
spacetime coordinates of points in spacetime with an operational meaning. Typically 
these come in terms of coordinate conditions, namely one specifies the coordinates of a point
by asking that four reference fields at the coordinates of that point take given values.
\item[2.] {\it Constraints, gauge transformations, Dirac observables}\\
In the Hamiltonian formulation that we are considering, the gauge symmetry of the theory is encoded 
in first class constraints. These generate gauge transformations on all fields by their 
Hamiltonian flow. A (weak) Dirac observable is a function of the fields which is invariant 
under that gauge flow when the constraints are satisfied.  
\item[3.] {\it Kinematical phase space, constraint surface, abstract reduced phase space}\\
While elegant, the definition of a Dirac observable is not constructive but abstract. 
One may distinguish the full, kinematical phase space of fields, the constraint surface 
where the constraints hold and the abstract reduced phase space which consists of equivalence classes
of points on the constraint surface which are related by gauge transformations. The latter 
is not a surface within the constraint surface but a moduli space of gauge orbits within 
the constraint surface. It is not explicit unless one can compute the gauge orbits concretely.
A Dirac observable can be considered as a function on that abstract reduced phase space as it 
depends only on the gauge orbits. Given a separating set of Dirac observables (i.e. they 
separate the points of the moduli space), they serve to equip the abstract reduced phase space 
with a Poisson bracket. This is simply the Poisson bracket on the kinematical phase space 
restricted to the Dirac observables which form a Poisson subalgebra of functions on the 
kinematical phase space due to the Jacobi identity.
\item[4.] {\it Relational observables, gauge fields, true degrees of freedom, gauge cut}\\
Relational observables are particular separating sets of Dirac observables which 
have the advantage that they come with a construction algorithm. While the general Dirac observables
can be referred to as abstract Dirac observables, the relational observables can be called concrete
Dirac observables. Their disadvantage is that 
they require additional input in the form of a choice of reference fields and gauge conditions
on them. In that context one splits the kinematical phase space into two disjoint sets of canonical pairs.
In the first set we find the reference fields and their conjugate momenta. This first set 
coordinatises the gauge degrees of freedom while the second set coordinatises the true degrees 
of freedom. Note that the notion of gauge and true degrees of freedom depends on the choice 
of reference fields. Then a relational observable subordinate to a choice of reference field and 
gauge condition on it is obtained by considering a function of the true degrees of freedom and its 
gauge transform with respect to that gauge transformation which installs the gauge condition {\it from anywhere} 
in the constraint surface. It can be considered as that gauge invariant function on the full 
constraint surface which when restricted to the gauge cut of the constraint surface defined by the gauge condition
returns the given function of the true degrees of freedom. Thereby points in the gauge cut    
can be considered as the representatives of the abstract gauge orbits. 
\item[5.] {\it Constraint surface branches, global issues, relational reduced phase space, concrete reduced phase space}\\
An important point in the concrete construction of the relational observables is that 
it requires us to solve the constraints explicitly for the gauge momenta conjugate to the reference fields. 
When the constraints depend non-linearly on those momenta, the constraint surface splits into
disjoint (up to lower dimensional boundaries) branches 
corresponding to choices of (square) roots. The gauge flow stops at the boundaries of the 
branches and thus one obtains relational Dirac observables for each choice of branch. 
We assume that they can be extended across those branches e.g. by continuity. This explicit 
form is important in order to explicitly compute
the Poisson algebra of those relational observables and to show that it is Poisson isomorphic 
with the subalgebra of functions on the kinematical phase space which depend only on the true 
degrees of freedom. We call the phase space coordinatised by the relational 
observables the relational reduced phase space, the phase space coordinatised by the 
the true degrees of freedom the concrete reduced phase space and functions thereon true observables.
\item[6.] {\it Physical Hamiltonian, reduced Hamiltonian, frozen picture}\\
Moreover, when at least one of the reference fields carries a gauge condition
that is explicitly time dependent, one obtains an explicit formula for the generator 
which drives this time evolution called the physical Hamiltonian.
This itself a relational observable of an, in general, explicitly time dependent function on the concrete 
reduced phase space called the reduced Hamiltonian. This time evolution of relational 
observables and, Poisson isomorphically, functions on the phase space of true degrees of freedom, 
is in contrast to the abstract Dirac observables which do not carry any time dependence. 
\item[7.] {\it Abstract Dirac versus relational observables}\\
1. Interpretation:\\
The time dependence of relational observables is of course introduced by hand.
It expresses the fact that if one fixes the gauge up to a one parameter freedom, the gauge fixing is 
incomplete and the remaining possible gauge flow that comes with changing that parameter 
becomes dynamics. This appears artificial but it is well motivated because its origin 
are the reference fields and these encode a reference frame. Since to directly compare theory calculations 
with experiment we must match the laboratory frame to the computational frame, the relational 
observables have the advantage that they have a direct physical interpretation while for the 
abstract Dirac observables such an interpretation is not directly available.\\
2. Quantisation·\\
When it comes to quantisation, the Poisson isomorphism between the Poisson algebra of 
relational observables and the Poisson algebra of functions of the true degrees of freedom
becomes particularly important. This is because it means that the relational observables 
associated to the position and momentum functions on the phase space of true degrees of freedom 
remain canonically conjugate and provide a separating and algebraically independent set of functions
of the the concrete 
reduced phase space. Therefore, it is straightforward to find Hilbert space representations
of the corresponding Weyl algebra. 

By contrast, given a set of abstract Dirac observables which 
we think of (weakly) gauge invariant functions on the kinematical phase space modulo those that 
vanish on the constraint surface that one has found by some independent procedure, these are typically 
highly non-linear and it is 
hard to decide whether they are separating and/or algebraically independent. 
Supposing that one has succeeded in identifying a separating set, their Poisson algebra closes 
but typically not with structure constants but with structure functions. Then it is 
not possible to use Lie algebra techniques to quantise this algebra, see \cite{10} for an 
example where this is still possible for constraints quadratic in the kinematical coordinate 
functions. Finally, the separating 
set may not be algebraically independent and finding an independent one may not even be 
possible because it would amount to solving algebraic equations of degree higher than four.
Then one may be able to resort to elaborate group theory quantisation techniques \cite{11} but 
the task of finding Hilbert space representations is now certainly not straightforward. On the 
other hand, the advantage of such an algebraically dependent set of abstract Dirac observables 
has the advantage of being free from the complications that comes with the branches.
\item[8.] {\it Quantisation before versus after reduction}\\
Quantising before reduction is in principle also straightforward because it is easy 
to construct Hilbert space representations of the coordinate functions of the entire 
kinematical phase space. However, now one faces several challenges. \\
1. One must find quantisations of the constraints.
This involves operator ordering issues and may lead to anomalies in the quantum constraint 
algebra because the classical constraint Poisson algebra is typically not a Lie algebra.\\
2. Even if those issues can be resolved one then must construct solutions of the quantum constraints 
as generalised zero eigenvectors of the kinematical Hilbert space which 
is not at all easy for non-linear constraints with structure functions because one cannot use 
rigged Hilbert space (group averaging) techniques.\\
3. Unless one finds the general 
solution by independent techniques, which is probably practically impossible 
in interacting quantum field theory such as quantum gravity,
one cannot start building a Hilbert structure (i.e. a new scalar product)
of those distributional solutions called the physical Hilbert space again because rigging techniques are not 
available. \\
4. A strategy to find the physical Hilbert space structure is then by asking that it carries a $^\ast$ representation 
of the algebra of abstract Dirac observables that preserve the space of these solutions.
This involves not only the classical complications 
of item [7.] for the process of actually finding that algebra
but also possible new anomalies in the quantum representation of that algebra may arise.

All the complications of 1.-4. are avoided when reducing before quantisation using relational 
observables because we directly and straightforwardly find Hilbert space representations of the 
$^\ast$ algebra of relational observables. The "only" non-trivial task left in that 
approach is to select those Hilbert space representations with respect to which its physical 
Hamiltonian becomes a self-adjoint operator and to deal with the existence of the branches which 
is tremendously simpler. 
\item[9.] {\it True degrees of freedom, observable map and gauge invariance}\\
As stated, the Poisson algebra of relational observables is isomorphic with the Poisson algebra 
of functions of the true degrees of freedom, i.e. at any given time 
defined by a reference frame, there exists a canonical transformation between 
them which is called the observable map. The physical Hamiltonian driving the dynamics of the 
relational observables is the image of the reduced Hamiltonian driving the dynamics of the 
true observables under this map. The observable map applied to a true observable can be written explicitly as an infinite series 
of iterated Poisson brackets between constraints and true observables. However, one never needs 
that series when one computes Poisson brackets between relational observables, all one needs 
to know is the Poisson bracket between true observables. When quantising, we quantise 
directly the algebra of relational observables and physical Hamiltonian which however is 
mathematically identical to quantising the true observables and reduced Hamiltonian. 
One therefore often does not distinguish between the two. This sometimes raises the criticism 
that the description in terms of the true degrees of freedom is gauge dependent because 
the true observables do not have (weakly) vanishing Poisson brackets with the constraints.
However, conceptually one needs to keep in mind is that the true observables are just 
a mathematical tool in order quantise the relational observables. Therefore there is 
no lack of gauge invariance because the latter are (weakly) gauge invariant by construction.  
The observable map is in fact defined on the full kinematical phase space and fails 
to be a canonical transformation between Poisson algebras of relational observables and functions 
on the kinematical phase space respectively. Rather, restricted to 
the constraint surface it is a projector that projects onto the Poisson algebra of relational 
observables. 
\item[10.] {\it Reference frames, relational observables and observational data}\\
An experiment can only measure what is observable and the basic assumption is that mathematically 
experimental observables are described by gauge invariant functions respectively 
operators in the classical respectively quantum theory. As the relational observables 
corresponding to coordinate functions of the true degrees of freedom 
provide a separating set, any observable can be written as a function of those. 
It therefore must be true, that any quantity that an experiment has access to must correspond to 
a relational observable. Now e.g. in General Relativity we can measure all ten components of the 
spacetime metric and as functions on the kinematical phase space these are certainly not 
gauge invariant. This apparent contradiction is resolved as follows: Assuming that 
the laboratory frame and the reference frame that was encoded by the choice of reference fields and 
gauge fixing conditions thereon coincide, we replace the 
kinematical metric by its restriction to the constraint surface and its projection 
by the observable map. In case that there are, as described in this paper, both primary and 
secondary first class constraints, the reference fields are used to fix the gauge freedom 
associated with the secondary constraints while the resulting gauge stability conditions 
fix the gauge freedom associated with the primary constraints. Together with the constraints, in this way, 
all functions on the kinematical phase space are written as functions of the true observables. We
then replace the true observables by their Poisson isomorphic relational observable thereby 
obtaining the same function but of the corresponding relational observables.
This is motivated by this isomorphism and the fact that when we evaluate those functions on the gauge cut defined 
by the gauge fixing condition, the two objects coincide. Therefore the prescription to construct a relational
observable from an arbitrary function on the kinematical phase space is simply given by 
using constraints, gauge fixing conditions and stability conditions to write it as a function 
of true observables only and then use the observable map. Its dynamics is then encoded 
by the physical Hamiltonian. 
\item[11.] {\it Relational reference frame dependence and relational reference frame transformation}\\
As the exposition reveals, the entire description of the gauge system at hand in terms of 
relational observables heavily 
relies on the choice of reference fields and gauge conditions thereon that in fact 
defines them. We have already 
explained that for any such choice, the description is gauge invariant because the extra structure 
that comes with those 
choices is employed to actually construct manifestly gauge invariant functions. Thus 
the dependence on the reference fields and gauge conditions is not in conflict with 
gauge invariance but rather expresses a dependence of the description on the reference 
frame. This of course triggers the question of how the description with respect to 
different reference frames are related to each other. We have explained in this 
paper that there exists, at the classical level, a natural, generically time dependent, canonical transformation 
between the Poisson algebras of relational observables subordinate to different reference 
frames. The existence of a bijection between these algebras follows from the fact that 
the number of algebraically independent relational observables cannot depend on the
reference frame. That it is even a canonical transformation follows from the finer 
details of the observable map: Applied to functions on the full constraint surface of the 
phase space it is not only a projector but also a Poisson homomorphism between 
the algebra of functions of relational observables and kinematical functions on the 
constraint surface equipped with the Dirac bracket defined by constraints and reference 
fields. However, the Dirac bracket by construction coincides with the Poisson bracket 
when applied to gauge invariant functions. Thus we may apply the observable map of one 
frame at a given time of that frame, to the images of the observable map of another frame at another time 
in that frame, to obtain a Poisson 
isomorphism between the relational observables labelled by those two different times. We call this map the 
relational reference frame 
transformation.
\item[12.] {\it Physical interpretation of the relational reference frame transformation}\\
The relational reference frame transformation is a canonical transformation between the 
relational observables defined by two different reference frames. As such it is not a gauge 
transformation but a transformation between Dirac observables, hence it should be assigned 
the status of a symmetry transformation. Since a relational reference frame is defined by a 
choice of a pair of reference fields and gauge conditions thereon, this symmetry transformation 
answers the question how to write the relational observables in one frame in terms of that of another
where the two frames can differ either 1. by only changing the gauge conditions on the same 
choice of reference fields or 2. by changing both reference fields and gauge conditions (changing 
just the reference fields but not the gauge conditions is generically ill defined as different reference fields 
generically cannot be subjected to the same gauge conditions, e.g. for dimensional reasons).

In case 1. such transformations can be interpreted as active rather than passive diffeomorphisms
(the latter are gauge transformations). The field theory case studied in the appendix provides an illustrative 
example and displays Lorentz transformations as such. 

In case 2. the change between reference frames is 
more drastic. An example for this would be the two mechanical models considered in 
the main text. An example from field theory would be the weak interaction where the role 
of the secondary constraints is played by the isospin Gauss constraint. The most familiar 
relational reference frame is given by choosing three out of four real Higgs fields as 
reference fields and their vanishing as gauge condition, known as unitary gauge. 
(Note that in this case the gauge condition is not explicitly time dependent which is 
possible because we must still fix the diffeomorphism gauge freedom by other reference fields 
where then the explicit time dependence resides). The Gauss 
constraint is solved for their conjugate momenta. The true degrees of freedom in the bosonic
sector are the respective three components of three real vector bosons and one real Higgs field as 
well as their conjugate momenta. Instead, 
one can pick one component out of three from each of the tree vector bosons as reference fields with 
a gauge condition on them, e.g. these could be longitudinal components with respect 
to some background metric and their vanishing as a gauge condition. One would then solve the Gauss      
constraint for their conjugate momenta. The true degrees are now all four real Higgs fields and the 
remaining two components each of the three vector bosons as well as their conjugate momenta. 
The relational reference frame transformation now expresses the relational observables corresponding 
to true degrees of freedom of the first frame in terms of the relational observables corresponding 
to true degrees of freedom of the second frame. In particular, when above gauge choices are 
taken, there is a canonical transformation 
which expresses the relational observables corresponding to four Higgs fields in the second frame 
in terms of the relational observables corresponding to all three components of all three vector 
bosons (and other fields) of the first frame. Upon quantisation,
the two descriptions have completely different 
Feynman diagrammes because the physical Hamiltonians depend on entirely different sets 
of independent fields, e.g. there are less interaction vertices between vector bosons 
in the second frame than there are in the first. Yet, when the corresponding quantum relational reference
frame transformation is unitary, then the scattering matrices are related by that unitary transformation. 
From a practical viewpoint it is amore convenient to use the first reference frame as 
one can solve the Gauss constraint algebraically rather than solving partial differential equations. 
On the other hand, for the electromagnetic interaction one picks the analog of the second reference frame 
and solves the electromagnetic Gauss constraint for the longitudinal part of the electric field 
rather than one of the electrically charged fermion modes.     
\item[13.] {\it Relational reference frame transformation and physical Hamiltonians}\\
Another question triggered by the reference frame dependence concerns the physical Hamiltonians.
The first guess might be that the physical Hamiltonians associated to the two frames are 
pullbacks of each other under the relational reference frame transformation. However, this 
is very generically not the case. The reason for this are manifold. First of all there 
maybe dimensional reasons. The choice of reference field is rather general and one may 
use fields of different (mass) dimension. The Hamiltonians have dimension conjugate to that 
of the reference field. Next there maybe range reasons. In different reference frames the Hamiltonians
may have different lower bounds. Furthermore, as the relational reference frame transformation 
is generically an explicitly time dependent canonical transformation, one may expect 
that the physical Hamiltonians differ by the effect of that time dependence on general grounds.
When the time dependence of the gauge condition is linear, this can lead to different 
physical Hamiltonians despite them being conservative, i.e. not explicitly time dependent 
with respect to the time defined by their reference frame. An illustrative example 
from field theory is discussed in the appendix. It shows  
that even for free scalar theories in Minkowski space the Hamiltonian is not a scalar under 
change of inertial reference frame, namely Poincar\'e transformations, but rather a 
component of the energy momentum 4-vector. It is translation and rotation invariant
but not boost invariant. To see how this comes about, consider the solution of the 
free Klein-Gordon equation with initial data on a fixed time hypersurface 
in an inertial frame. Consider another inertial frame relatively boosted to 
the first and a fixed time hypersurface in that frame.
Then take the time derivative of the solution with respect to both notions of time. 
The fields and their time derivatives restricted to either constant time hypersurface
satisfy the canonical commutation relations. Thus for each fixed time in both frames 
we obtain a corresponding canonical transformation. Now consider the Hamiltonian 
in the second frame. As the system is conservative, we may write it in terms of fixed 
time field and time derivative of the solution. Using the just established canonical 
transformation we may also write it in terms of the initial data of that solution with respect
to a constant time hypersurface in the first frame. However, that does not immediately yield
the Hamiltonian in the first frame on that constant time hypersurface of the first frame 
because the solution that enters the second Hamiltonian is not spatially integrated on a constant time hypersurface
of the first but rather the second frame. In order to obtain an integral over the a constant time 
hypersurface of the first frame, we must undo the boost that enters the solution by conjugating 
the Hamiltonian of the first frame by a boost generator. The result is a linear combination of 
Hamiltonian and momentum in boost direction with respect to the first frame. 
In the appendix this process is embedded into the general constraint system framework of the main text.

Thus, the relational frame transformation and the physical Hamiltonians in the the corresponding two 
frames are correctly interrelated: Taking two reference frames and their corresponding 
physical Hamiltonians, we may use the relational reference frame transformation from the first to 
the second frame to pullback the physical Hamiltonian of the second frame. The result is a Hamiltonian in 
the first frame, in particular it is a relational observable, 
but it is not necessarily the physical Hamiltonian of that frame. Generically, also
for general gauge systems, we expect that it contains a correction 
corresponding at least to the analog of the boost part of the relational reference frame transformation
interpreted as an active diffeomorphism of spacetime. Only when one uses 
the physical Hamiltonian determined by a reference frame to derive the dynamics relational observables determined 
by the same reference frame, does one obtain the correct dynamical description of the system. 
\item[14.] {\it Matching laboratory and computational reference frame}\\
When using a computational reference frame associated with a choice of reference 
fields and gauge fixing conditions and considering any kinematical field of the theory then we use the
observable map of that frame to obtain an observable associated to that field as described 
in item[11.]. This in particular means that the reference fields themselves are replaced by 
the corresponding gauge condition values times the unit function (or unit operator in the 
quantum theory). However, the experimentalist may find different values from those. 
If the theory is correct, this simply means that the computational frame and the laboratory frame are not aligned.
Hence, either the he theorist or the experimentalist 
has to adapt (to) the calculation. The theorist can do this either by 
monitoring experimentally the time evolution of the chosen or other reference fields  
and using those as gauge fixing conditions and reference fields as input for the calculation or by recycling the computation already 
done in the misaligned computational frame and using the relational reference frame transformation
corresponding to different gauge conditions on the same or different reference fields. 
Given the generically quite complicated form of the relational reference frame transformation,
the first option seems to be preferred. The experimentalist can deduce the spacetime diffeomorphism
that mediates between the measured and gauge fixed values of the same reference fields 
and use it to pull back the experimental data by that diffeomorphism. 

At the classical 
level and given infinite measurement precision these two procedures are equivalent. 
At the quantum level, this is not obvious: The quantum relational reference frame transformation 
is between operators while translating measurement values is translating between expectation values
of those operators that are found by many, non destructive, repetitive measurements of the metric.  
\item[15.] {\it Quantum relational reference frame transformation and fluctuations}\\
When quantising the system the relational reference frame transformation also gets quantised and 
given that it is classically a canonical transformation, it an interesting question whether one can  
turn it into a unitary transformation. Especially in Quantum Field Theory we expect this
to be a hard problem given the fact that even generic Bogol'ubov transformations between 
free quantum fields fail to be unitarily implementable.     

Another aspect of this is the confusing fact that the reference field used to define a 
reference frame is always quantised
as the value given by the corresponding gauge fixing condition times the unit operator.
Thus given a second, different reference frame in which the first reference field is considered a true 
degree of freedom, it gets quantised as a non-unit operator. The first quantisation 
yields zero fluctuations, the other non-zero fluctuations with respect to generic states.
The resolution of that paradox as we explained in the main text is that the two 
operators are in fact different quantum observables because when switching reference frames 
observables are non-trivially mapped among each other. The correct translation of 
the first observable to the second reference frame or the second observable to the first is 
by using the quantum relational reference frame transformation. If it is unitary, the unit operator remains the 
unit operator and displays zero fluctuations also in the second frame. Likewise unitary 
operations cannot turn a non-unit operator into the unit operator. 
\item[16.] {\it Quantisation before or after reduction and quantum clocks}\\
Yet another confusing notion is that of quantum clocks and the role that reference 
fields play in this. We have seen that the relational observables corresponding 
to the reference fields defining a reference frame 
are quantised as proportional to the unit operator in the reduction before quantisation 
scheme. On the other hand, one may consider quantisation before reduction and thus the 
reference fields become non-trivial operators on the kinematical Hilbert space with 
non-zero fluctuations. The puzzle is how that can be true. A consistent viewpoint is 
as follows and rests on the basic assumption that all we can ever  measure are 
gauge invariant objects: In the reduction after quantisation approach we should   
therefore ask whether the quantised reference field is observable, i.e. gauge invariant.
{\it It is not!}. This is in contrast to the relational observable corresponding to 
the reference field which is trivially gauge invariant (by definition and explicitly as 
it is a multiple of the unit operator).

We could ask a similar question in quantum electrodynamics on Minkowski 
space: Can we measure the vector potential (say smeared over an arbitrarily small
spacetime region)? We cannot, because that is not a gauge invariant object. What we can measure are 
$U(1)$ gauge invariant objects like electric and magnetic fields or the line integral of the 
vector potential for closed paths not bounding a 2-surface (knots; if the closed path 
is a boundary, we can use the (distributional) Stokes theorem to obtain an expression
in terms of the (distributional) Faraday tensor). Therefore, while the 
reference field becomes a non-trivial operator (valued distribution) in the quantisation
before reduction scheme, we can still only measure gauge invariant objects constructed 
from this, like relational observables. 

There is of course the issue whether the two quantisation schemes commute and in general 
they will not. In the reduction before quantisation scheme, one directly 
quantises the Poisson algebra of relational observables. In the quantisation before reduction 
scheme one would consider the infinite series expression that writes the relational 
observable in terms of reference fields and true degree of freedom fields, choosing 
some operator ordering of all those terms and study that operator on the Hilbert space
defined by the zero eigenvector solutions of the quantum constraints which are 
also written in terms of reference and true fields. The two schemes should have the same 
semiclassical limit but one would expect the quantum corrections to be scheme dependent
in general, see \cite{13}. However, this an instant of the general fact that there is 
no unique path to quantisation of a classical system rather than wrongly assigning a gauge 
non-invariant object the status of an observable.

Rather, an observable quantum clock could be the quantisation of any relational observable
such as an actual atomic clock in terms of quantum Caesium atoms and the quantised electromagnetic 
field (which involves all relational observables corresponding to 
standard model fields for a full description). Thus one enters here the realm of quantum physics 
in which the apparatus is considered as part of the quantum system rather than approximating 
it by an outside 
classical observer which is also the context in which quantum clocks are mostly 
discussed in the theory of quantum reference frames. But again we are talking only about 
observables, i.e. gauge invariant objects. 
\item[17.] {\it Explicit time dependence}\\
Throughout this work we have imposed gauge fixing conditions which are explicitly 
time dependent for at least one of the modes of the reference fields. For example 
we may select four components of the spacetime metric tensor and and impose that 
one of them depends only on time while the other three are time independent and instead 
take certain numerical values. The four numerical values given by these specify 
the spacetime point when the reference field is known. 
Such gauge fixing conditions are appropriate when the Cauchy slices of the 
globally hyperbolic spacetime have no boundary. By contrast, in asymptotically flat  
situations it is possible to drop the explicit time dependence in the gauge fixing conditions
because in this case the constraints have to be augmented by boundary terms. 
Now it is those boundary terms that become the physical Hamiltonian which is now 
conservative. Technically, the stability condition associated with the gauge fixing 
condition which involves solving partial differential equations contains 
integration constants and thus display incomplete or equivalently a one parameter 
family of gauge fixings quite similar to the mechanism by which the explicit time parameter family does. 

In the case without boundary, those integration constants can be gauged away. However, 
in the case of a boundary there are boundary conditions, e.g. that the metric is asymptotically flat 
(Minkowski), in the form of decay behaviour 
of the fields at spatial infinity which do not allow these integration constants to be gauged away.
The form of the boundary term is also dictated by these decay conditions. 

Now consider two reference frames specified by two different choices of reference fields 
and/or gauge fixing conditions. Then one must adapt the decay of all the fields 
to the choice of these gauge fixing conditions \cite{12,14} in order to comply with asymptotic 
flatness. This changes the reference frame and potentially the boundary term and thus the physical 
Hamiltonian. A change of reference frame is a symmetry transformation rather than a gauge transformation
precisely when it changes the boundary term. For example we may use gauge fixing conditions on the spatial part of 
the metric which impose among other things that the asymptotic coordinate system 
are Gullstrand-Painlev\'e coordinates rather than Schwarzschild for a given total mass of the system. 
Then the information about 
the mass of the system sits in the ADM momentum \cite{7} with respect to the first frame while it 
sits in the ADM energy with respect to the second \cite{15}. This is in accordance 
with the fact that the Gullstrand Painlev\'e foliation is (non-linearly) boosted with respect to the 
Cartesian foliation and it is well known that the ADM charges play the role of 
Energy-Momentum vector of the asymptotically Minkowskian spacetime \cite{12}. In that sense 
this particular change of reference frame is very similar to a Lorentz transformation 
discussed in the appendix. The way that non-linear Lorentz transformation is encoded in time independent 
gauge fixing conditions on reference fields of the secondary phase space is because they 
induce corresponding gauge fixing conditions on the primary phase space, that is, lapse and 
shift functions. These encode the temporal-temporal and temporal-spatial components 
of the spacetime metric with respect to a foliation and thus can only change when the 
corresponding spacetime diffeomorphism has a non-trivial time dependence.    

Gravitational boundary term analyses in the context of quantum reference frames were also recently considered 
in \cite{15a}. 
\item[18.] {\it Quantum relational reference frames and other notions of quantum reference frames}\\
As we have outlined at the end of section ref{s7}, a possible link between quantum relational 
reference frames and the bulk of the work on quantum reference frames is that general quantum relational 
reference frame transformations could, at least partly, be the restriction to selected degrees of freedom of 
quantum relational reference transformations between relational reference frames defined by 
different reference fields and/or gauge fixing conditions. Indeed, in the bulk of research on
quantum reference frames one considers the absence of any gauge redundancy, therefore 
such a link necessarily must involve relational observables. The interesting point is now that 
quantum relational reference frame transformations involve strong dynamical imprint from 
the underlying microscopic quantum gauge field theory and are not extra kinematical structure
that one can choose independent of the dynamics. 
\end{itemize}
This long list of items displays how a concrete quantisation scheme for 
generally covariant quantum fields automatically provides an interface with the fascinating 
topic of quantum reference frames through the relational reference frame transformation which 
is induced by the relational observable projection maps given by two pairs of reference fields 
and gauge fixing conditions. Here we have mostly outlined the concepts and provided a few 
non-trivial examples but worked them out mostly  in the classical theory (see however 
the appendix). Also we only mentioned open technical problems in particular when it comes to 
quantisation in section \ref{s7} but did not solve them. It is mandatory  
to work out many more and more complicated examples and quantise the corresponding relational reference frame 
transformation to gain further insight into these and related questions.

\begin{appendix}

\section{Hamiltonians and reference frames for scalar fields in Minkowski spacetime}
\label{sa}

The present section describes the dependence of Hamiltonians on the inertial frame for 
scalar fields in Minkowski space. To connect it to the main formalism we display those theories 
in terms of parametrised field theories \cite{16} which are generally covariant versions 
of those theories by considering the spacetime coordinates as dynamical fields, namely as 
dynamical diffeomorphisms. 
We will see that the relational reference frame transformation is here not due to a switch between
gauge and true degrees of freedom but rather due to a switch of gauge conditions on the same 
reference fields which here are chosen as those dynamical coordinates.
A choice of gauge condition on the reference fields is then nothing but a choice of spacetime 
foliation. 

In the first subsection we review the parametrised field theory framework. 
In the second we apply this formalism to the relational reference frame transformation
between arbitrary foliations. In the third subsection we 
consider the special case of a switch between gauge fixing conditions that encode
inertial frames. 

We can do this in any spacetime dimension and in the classical theory 
for arbitrary (polynomial) self-interaction potentials. In the quantum theory, these constructions 
translate literally and rigorously only for quadratic potentials in corresponding Fock 
representations at the operator level, with self-interactions these considerations are  
still valid perturbatively at the level of quadratic forms on the Fock space 
of the free part and with proper renormalisation understood. We note that applied 
to the case of uniformly accelerated frame gauge fixing conditions we recover the 
Hamiltonian treatment of the Unruh effect \cite{17}. As we will see, nothing really depends 
on the type of theory, we could have considered any other Poincar\'e covariant Lagrangian,
we just chose scalar fields for illustrative purposes.

\subsection{Review of parametrised field theory}
\label{sa.1}

We consider the spacetime manifold $M=\mathbb{R}^{D+1}$ with Minkowski metric $\eta$ and 
global Cartesian coordinates $x^A,\;A=0,1,..,D$.  
The Lagrangian of the (self-interacting, real) scalar field on Minkowski spacetime is given by 
\be \label{a.1}
L=-\frac{1}{2}[\eta^{AB}\;\Phi_{,A}\;\Phi_{,B}+V(\Phi)]
\ee
for some polynomial potential $V$. The Lagrangian is of course Minkowski background dependent. 

We now consider $x^A=x^A(z)$ as a scalar field by itself where $z^\mu,\; \mu=0,..,D$ are also 
Cartesian coordinates in $\mathbb{R}^{D+1}$ subject to the condition that $\det(\partial x/\partial z)>0$,
i.e. $z\mapsto x(z)$ is an orientation preserving diffeomorphism of $\mathbb{R}^{D+1}$. We introduce 
the pullback fields $\Phi'=x^\ast \Phi,; g=x^\ast \eta$, explicitly 
\be \label{a.2}  
\Phi'(z)=\Phi(x(z)),\; g_{\mu\nu}(z)=x^A_{,\mu}(z)\;x^B_{,\nu}(z)\eta_{AB}
\ee
These are considered fundamental from now on while $\Phi$ is considered a derived concept obtained 
from $\Phi'(z)$ by inverting the diffeomorphism.
With those we build the parametrised field theory (PFT) Lagrangian
\be \label{a.3}
L'=-\frac{1}{2}|\det(g)|^{1/2}\; [g^{\mu\nu}\;\Phi'_{,\mu}\;\Phi'_{,\nu}+V(\Phi')]
\ee
It is not difficult to check that $L=L'$ when $x$ is the identity diffeomorphism. 
On the other hand, (\ref{a.3}) is a scalar density under arbitrary diffeomorphisms 
(coordinate transformations) of $z$, called reparametrisations. One often distinguishes 
between $x$ space as target space and $z$ space as parameter space. Note that $L'$ depends 
on the fields $\Phi',x^A$ and (their first order partial $z$ derivatives) as independent 
Lagrangian degrees of freedom. 

The Hamiltonian treatment of $L'$, using the $z^0=$const. hypersurfaces as foliation 
in $z$ space asks to compute the momenta $\pi:=L'_{,\Phi'_{,z^0}}, \;
y_A:=L'_{,x^A_{,z^0}}$ at $z^0=0$ and invert those for the velocities $U:=\Phi'_{,z^0},
X^A:=x^A_{,z^0}$ and to define the primary Hamiltonian as 
$H'=U\pi+X^A \;y_A-L$ at the solution of those inversions. The computation is 
standard \cite{16}. It turns out that expectedly one cannot solve for any of the 
$X^A$ and one finds 
\be \label{a.4}
H'=X^A\; C_A,\; C_A:=y_A+h_A,\; h_A:=n_A \; Z+q^{ab}\eta_{AB} X^B_{,a} \;Z_b,\;
Z_a=\pi\phi_{,a},\;
Z=\frac{1}{2}(\frac{\pi^2}{\det(q)}+q^{ab}\phi_{,a}\phi_{,b}+V(\phi))
\ee
Here $q_{ab}=g_{ab};\; a,b,c,..=1,..,D, \; q^{ac} q_{cb}=\delta^a_b$. Note that all 
quantities are evaluated at $z^0=0$, in particular we denote $\phi$ as the restriction
of $\Phi'$ to the $z^0=0$ hypersurface. An important role is played by the co-normal 
\be \label{a.5}
n_A=\frac{1}{D!}\epsilon_{A B_1..B^D}\epsilon^{b_1 .. b_D} \;x^{B_1}_{,b_1}..
\;x^{B_D}_{,b_D}
\ee
with convention $\epsilon_{01..D}=1=\epsilon^{12..D}$. We note the identities
\be \label{a.6}
\eta^{AB} n_A n_B=-\det(q),\; n_A\; x^A_{,a}=0
\ee
It follows that $H'$ is fully constrained with constraints equivalent to normal and tangential 
projections 
\be \label{a.7}
c:=\eta^{AB} y_A n_B-\det(q)\;Z, \;c_a=y_A\; x^A_{,a}+Z_a
\ee
These are first class, their Poisson brackets displaying an instant of the hypersurface 
deformation algebroid \cite{18} and are thus called Hamiltonian and spatial 
diffeomorphism constraint. Conversely, the primary constraints $C_A$ are the solution of (\ref{a.7}) for 
$y_A$ which have an Abelian constraint algebra. In the notation of the main text, we thus have $C=\bar{C}$ in 
this theory and the constraint surface has a single 
branch because the constraints depend linearly on $y$. In particular, there are no secondary constraints.

To fully embed the present formulation into the notation of the main text we 
artificially introduce momenta $Y_A$ conjugate to $X^A$ and consider the augmented Hamiltonian
\be \label{a.8}
H'=V^A\; Y_A+X^A \; C_A
\ee
with undetermined velocities $V^A$ of $X^A$ enforcing the primary constraints $Y_A=0$ whose 
stability enforces the secondary constraints $C_A=0$. Introducing some real valued orthonormal
basis $b_\alpha$ of $L_2(\mathbb{R}^D, d ^Dz)$ we consider the coefficients 
$\phi^\alpha:=<b_\alpha,\phi>, \;\pi_\alpha:=<b_\alpha,\pi>$ as representing the conjugate true degrees of freedom $q^a,p_a$ and 
$x_\alpha^A:=<b_\alpha,x^A>, \;y^\alpha_A:=<b_\alpha,y_A>$ as representing the conjugate gauge degrees of freedom $x^I,y_I$
and similar for $X^I, Y_I$.

\subsection{Changing between general coordinate gauge conditions}
\label{sa.2}

As mentioned, we will not consider switching between $(q,p)$ and $(x,y)$ to define different sets of relational 
observables, rather we will consider switches of the gauge condition on the same choice of reference fields 
$x,y$ which here takes the form
\be \label{a.9}
G^A(z):=x^A-k^A(z)
\ee 
where $k^A(z)$ are coordinate conditions. This is precisely of the required form upon decomposing with
respect to the basis $b_\alpha$ and the interpretation $I=(\alpha,A)$ as compound label and the time label $z^0=t$
\be \label{a.10}
G^I(t):=<b_\alpha, G^A(z^0=t,.)>=x^A_\alpha-k^A_\alpha(t)
\ee 
The adaption of the PFT formulation to the main text enables us to immediately write the reduced 
Hamiltonian as
\be \label{a.11}
h=\dot{k}^I(t)\;h_I(x=k(t),q,p)=\sum_\alpha\; \dot{k}^A_\alpha(t) \;<b_\alpha,h_A>=
\int\; d^Dz\; \dot{k}^A(t, \vec{z})\; h_A(x=k(t,\vec{z})),\phi(\vec{z}),\pi(\vec{z}))
\ee
where the completeness relation was used. We use here the notation 
$z=(z^0,\vec{z})$ to distinguish between parameter time and parameter space coordinates. 

A separating set of corresponding relational observables is given by 
\be \label{a.12}
\Phi'(t,\vec{z}):=O_{\phi(\vec{z})}(t), \Pi'(t,\vec{z}):=O_{\pi(\vec{z})}(t), \; 
O_F(t)=[\exp(\{<g^A,C_A>,.\})\cdot F]_{g=-G}
\ee 
As emphasised throughout the main text, these of course depend explicitly on $k$ but are conjugate 
$\{\Pi(t,\vec{z}_1),\Phi(t,\vec{z}_2)\}=\delta(\vec{z}_1,\vec{z}_2)$.

A set of gauge fixing conditions defines a foliation of target space by constant $t$ hypersurfaces 
\be \label{a.13}
\Sigma_t:=\{x^A=k^A(t,\vec{z});\; \vec{z}\in\mathbb{R}^D\} 
\ee
Let $\Phi'(\vec{z}):=\Phi'(0,\vec{z}),\; \Pi'(\vec{z}):=\Pi'(0,\vec{z})$ be the relational 
conjugate coordinates on $\Sigma_0$. Then the relational conjugate coordinates 
$\Phi'(t,\vec{z}),\;\Pi'(t,\vec{z})$ are the time evolutions with respect 
to the physical Hamiltonian
\be \label{a.14}
H(t)=\int\; d^Dz\; \dot{k}^A(t, \vec{z})\; h_A(x=k(t,\vec{z})),\Phi(\vec{z}),\Pi(\vec{z}))
\ee
i.e. 
\be \label{a.15}
\Phi'(t,\vec{z})={\cal P}_r\; \exp(\int_0^t\; ds\; \{H(s),.\})\cdot \Phi'(\vec{z}) 
\ee
and similar for $\Pi'(t,\vec{z})$. 

Consider now a second foliation defined by a different gauge fixing function
$\hat{k}^A(z)$. We denote the corresponding objects defined by this second 
foliation by a hat. The relational reference frame transformation, labelled by $t,\hat{t}$ now consists in expressing the
fields defined by the second foliation on a hypersurface $\hat{\Sigma}_{\hat{t}}$ in terms
of the fields defined by the first foliation on a hypersurface $\Sigma_t$. This is 
accomplished by 
\be \label{a.16}
\hat{\Phi}'(\hat{t},\vec{z})=O_{\hat{\Phi}'(\hat{t},\vec{z})}(t)=
O_{\hat{O}_{\phi(\vec{z})}(\hat{t})}(t)
\ee
and similar for $\hat{\Pi}'(\hat{t},\vec{z})$. In the present case, we do not need to use the 
constraints to write the right hand side of (\ref{a.16}) in terms of $\Phi'(t,\vec{z}),\Pi'(t,\vec{z})$
because we do not 
switch between reference fields. Formula (\ref{a.16}) asks us to to write 
$\hat{O}_{\phi(\vec{z})}(\hat{t})$ explicitly as a function of $\phi(\vec{z}),\pi[\vec{z}],x^A(\vec{z})$ 
and then to apply $O_\cdot(t)$ to that expression which results in the same expression 
with the substitution of arguments by $\Phi'(t,\vec{z}),\Pi'(t,\vec{z}),k^A(t,\vec{z})$.
By the properties of the observable maps $O_\cdot(t),\hat{O}­\cdot(hat{t})$ this is a 
a canonical transformation.

\subsection{Changing between inertial frames}
\label{sa.3}

The discussion so far applies to any two pairs foliation and can be used to consider the Unruh effect 
or the question when such a transformation can be unitarily implemented  \cite{9}. In the case of 
free fields (\ref{a.16}) is then a Bogol'ubov transformation \cite{17}. We now specialise to 
the following two foliations
\be \label{a.19}
k^A(t,\vec{z}):=t\;\delta^A_0 +z^a\delta^A_a, \;
\hat{k}^A(\hat{t},\vec{z})=L^A_0\; \hat{t}+L^A_a \; z^a
\ee
where $L$ is a proper, orthochronous, non-trivial Lorentz transformation. The first gauge fixing condition 
means that the diffeomorphism between target and parameter space is the identity map, the second that 
it is a Lorentz transformation. In the latter case we are considering an example in which 
not only one of the $x^A$ develops an explicit time dependence which shows the necessity to allow 
for this flexibility in the gauge fixing conditions.

We construct the corresponding physical Hamiltonians.
In both cases we have $q_{ab}(\vec{z})=\delta_{ab}$ because $L$ is a Lorentz transformation 
$L^A_\mu L^B_\nu \eta_{AB}=\eta_{\mu\nu}$.  
For the trivial Lorentz transformation we have $\dot{k}^A=\delta^A_0,\;
n_A=\delta_A^0,\; X^A_{,a}=\delta^A_a$. For the non-trivial Lorentz transformation we have 
$\dot{k}^A=L^A_0,\;L^A_0 n_A=\det(L)=1,\; \eta_{AB} L^A_0 X^B_{,a}=\eta_{0a}=0$.  
Thus in both cases we obtain the not explicitly time dependent result
\be \label{a.20}
h=\hat{h}=\int\; d^D z\; Z=\frac{1}{2}\int\; d^D z\;(\pi^2+\delta^{ab}\phi_{,a}\phi_{,b}+V(\phi))
\ee
The corresponding physical Hamiltonians are therefore simply 
\be \label{a.21a}
H=\int\; d^D z\; Z=\frac{1}{2}\int\; d^D z\;(\Pi^2+\delta^{ab}\Phi_{,a}\Phi_{,b}+V(\Phi));,
\hat{H}=\int\; d^D z\; Z=\frac{1}{2}\int\; d^D z\;(\hat{\Pi}^2+\delta^{ab}\hat{\Phi}_{,a}\hat{\Phi}_{,b}+V(\hat{\Phi}));,
\ee

In the main text, we discussed two canonical transformations between relational observables,
for fixed $t,\hat{t}$. 
One is given by the trivial canonical transformation $\hat{\Phi}'(\hat{t},\vec{z}),\hat{\Pi}'(\hat{t},\vec{z})
=S(\Phi'(t,\vec{z}),\Pi'(t,\vec{z})):=()\Phi'(t,\vec{z}),\Pi'(t,\vec{z}))$ and the other by 
(\ref{a.16}). Under the trivial transformation we have in this case $\hat{H}\circ S=H$. However for the relative 
reference exchange map $S$ we must work out (\ref{a.16}). We could do this by going through the explicit 
series expansion of the observable maps. But this requires a rather non-trivial calculation because 
while $\phi,\pi$ do not explicitly depend on $x^A$, already the first Poisson bracket with $C_A$ does in a
very complicated way. We will therefore compute the map by indirect methods. 

Let $L$ be any proper orthochronous Lorentz transformation, be it trivial or not, $f$ any function 
depending only on the true degrees of freedom $\phi(\vec{z}),\pi(\vec{z})$ and not explicitly on $z^0$. 
For better readability 
we define $O^L_f(z^0):=[\exp(\{<g^A,C_A>,.\})\cdot f]_{g=Lz-x}$. The strategy will be to first 
express this just in terms of $O^L_f(0)$ and then to relate $O^L_f(0), O^{L'}_f(0)$ for different 
choices of $L,L'$ specifically $L'=$id. Although we could recycle some of the properties 
of the observable map from section \ref{s2}, it is instructive to perform the derivations from scratch
in the present case. We have
\be \label{a.21b}  
\frac{d}{dz^0}\; O^L_f(z^0)=O^L_{\{h_L,f\}}(z^0),\;h_L=\int\; d^D z\;L^A_0 h_A(\vec{z})
\ee
As $h_L,f$ do not depend on $y_A$ we have $\{h_L,f\}=\{h_L,f\}^\ast$ where the latter is the 
Dirac bracket for $C_A, G^A=x^A-L^A_\mu z^\mu$. Then 
\be \label{a.21}  
\frac{d}{dz^0}\; O^L_f(z^0)=\{O^L_{h_L}(z^0),O^L_f(z^0)\}
\ee
Now consider the explicit expression $h_L[x;\phi,\pi]$ as a functional of $x,\phi,\pi$. 
Applying $O^L_\cdot(z^0)$ to it replaces $x^A(\vec{z})$ by $L^A_\mu z^\mu$ and 
$\phi(\vec{z}),\pi(\vec{z})$ by $O^L_{\phi(\vec{z})}(z^0), O^L_{\pi(\vec{z})}(z^0)$. 
Now under this substitution
\be \label{a.22}
O^L_{x^A_{,a}(\vec{z})}(z^0)=L^A_a,\; O^L_{n_A(\vec{z})}(z^0)=\det(L) [L^{-1}]_A^0, \;O^L_{q_{ab}(\vec{z})}(z^0)=\delta_{ab}
\ee
whence  
\ba \label{a.23}
&& O^L_{h_L}(z^0)=\int\; d^D z\;L^A_0 
((L^{-1})_A^0\frac{1}{2}([O^L_{\pi(\vec{z})}(z^0)]^2+\delta^{ab}[O^L_{\phi(\vec{z})}(z^0)]_{,a}[O^L_{\phi(\vec{z})}(z^0)]_{,b}
+V([O^L_{\phi(\vec{z})}(z^0)])
\nonumber\\
&& +\eta_{AB} L^B_a\delta^{ab} [O^L_{\pi(\vec{z})}(z^0)][O^L_{\phi(\vec{z})}(z^0)]_{,b})
=O^L_h(z^0),\; h=\int\; d^D z\; \frac{1}{2}[\pi^2+\delta^{ab}\phi_{,a}\phi_{,b}+V(\phi)](\vec{z})
\ea
We note that $h$ is independent of $L, z^0,x^A$. Thus we me apply (\ref{a.21}) to $f=h$ resulting in 
\be \label{a.24}
\frac{d}{dz^0}\; O^L_h(z^0)=\{O^L_h(z^0),O^L_h(z^0)\}=0
\ee
Therefore 
\be \label{a.25}
H_L:=O^L_h(z^0)=O^L_h(0)=\int\; d^D z\;\frac{1}{2}(\pi^2+\delta^{ab}\phi_{,a}\phi_{,b}+V(\phi)](\vec{z}))_{\phi\to O^L_{\phi}(0),\pi\to O^L_\pi(0)}
\ee
i.e. the physical Hamiltonian is not explicitly time dependent and can be written in terms of the time zero 
relational observables. Then (\ref{a.21}) has the unique solution  
\be \label{a.26}
O^L_f(z^0)=\exp(z^0 \;\{H_L,.\})\cdot O^L_f(0)
\ee
and thus all functions of interest can be written in terms of the time zero relational observables corresponding 
to the reference frame determined by $L$.

To write $O^L_f(0)$ in terms of $O^1_f(0)$ where $O^1_f(0)$ are the relational observables at time zero 
for the identity Lorentz transformation, we consider a 1-parameter group of Lorentz transformations 
$[0,s_0]\to {\cal L}^\uparrow_+;\; s\mapsto e^{s\;M}$ where $M$ is a Lie algebra element such that 
$e^{s_0 M}=L$. Then we have again for a function only depending on $\phi,\pi$ but not on $x^A, s$ 
\be \label{a.27}
\frac{d}{ds}\; O^{L(s)}_f(0)=O^{L(s)}_{\{\kappa_L(s),f\}}(0),\;
\kappa_{L(s)}=\int\; d^Dz\;z^a\dot{L}^A_a(s)\;h_A(\vec{z})
\ee
The reasoning is now rather similar as for the time translations: As $\kappa_{L(s)},f$ do not depend on 
$y_A$ we have 
$\{\kappa_L(s),f\}=\{h_L,f\}^\ast$ so that 
\be \label{a.28}  
\frac{d}{ds}\; O^{L(s)}_f(0)=\{O^{L(s)}_{\kappa_{L(s)}}(0),O^{L(s)}_f(0)\}
\ee
Again consider the explicit expression $O^{L(s)}_{\kappa_{L(s)}}(0)[x;\phi,\pi]$ as a functional of $x,\phi,\pi$. We obtain
\ba \label{a.29}
&& O^{L(s)}_{\kappa_{L(s)}}(0)=\int\; d^D z\;\dot{L}^A_0(s) ((L(s)^{-1})_a^0\frac{1}{2}([O^{L(s)}_{\pi(\vec{z})}(0)]^2
+\delta^{ab}[O^{L(s)}_{\phi(\vec{z})}(0)]_{,a}[O^{L(s)}_{\pi(\vec{z})}(0)]_{,b}
+V([O^{L(s)}_{\phi(\vec{z})}(0)])
\nonumber\\
&& +\eta_{AB} L^B_b(s)\delta^{bc} [O^{L(s)}_{\pi(\vec{z})}(0)][O^{L(s)}_{\phi(\vec{z})}(0)]_{,c})
=O^{L(s)}_{\kappa_M}(0),\; 
\nonumber\\
\kappa_M &=&\int\; d^D z\; z^a\dot{L}^A_a(s)[(L(s)^{-1})^0_A\frac{1}{2}[\pi^2+\delta^{ab}\phi_{,a}\phi_{,b}+V(\phi)]
+\eta_{AB} L^B_b(s)\delta^{bc} \pi\phi_{,c}](\vec{z})
\ea
The notation suggests that $\kappa_M$ does not depend on $s$. This is indeed true since 
\be \label{a.30}
\dot{L}^A_a(s) (L(s)^{-1})_A^0=L^A_\mu(s) M^\mu_a (L(s)^{-1})^0_A=M^0_a,\;
\dot{L}^A_a(s) \eta_{AB} L^B_b(s)\delta^{bc}=L^A_\mu(s) M^\mu_a \eta_{AB} L^B_b(s) \delta^{bc}=\delta_{ac} M^c_d \delta^{db} 
\ee
Thus 
\be \label{a.31}
\kappa_M=\int\; d^D z\; z^a\;[M^0_a\frac{1}{2}[\pi^2+\delta^{ab}\phi_{,a}\phi_{,b}+V(\phi)]
+\delta_{ac} M^c_d \delta^{db}\pi\phi_{,c}](\vec{z})
\ee
Since $\kappa_M$ only depends on $\phi,\pi$ but not on $x^A,s$ we may apply (\ref{a.28}) to $f=\kappa_M$ and find
\be \label{a.32}
\frac{d}{ds}\; O^{L(s)}_{\kappa_M}(0)=\{O^{L(s)}_{\kappa_M}(0),O^{L(s)}_{\kappa_M}(0)\}=0
\ee
which means that 
\be \label{a.33}
K_M:=O^{L(s)}_{\kappa_M}(0)=O^{L(0)}_{\kappa_M}(0)=O^1_{\kappa_M}(0)=\kappa_M[\phi\to O^1_\phi(0),\pi\to O^1_\pi(0)]
\ee
which expresses $K_L$ as a functional of the time zero relational observables of the reference frame determined by $L=1$. Then 
\be \label{a.34}  
\frac{d}{ds}\; O^{L(s)}_f(0)=\{K_M,O^{L(s)}_f(0)\}
\ee
has the unique solution 
\be \label{a.36}
O^{L(s)}_f(0)=e^{s\{K_M,.\}}\cdot O^{L(0)}_f(0)=e^{s\{K_M,.\}}\cdot O^1_f(0)
\ee
Concluding 
\be \label{a.37}
O^L_f(0)=e^{s_0\{K_M,.\}}\cdot O^1_f(0)
\ee
Hence to express the relational observables of the time zero fields of the foliation determined by $L$ we 
evolve the time zero fields of the foliation determined by $L=1$ by the generator $K_M$ a parameter distance $s_0$
where $K_M$ is written explicitly in terms of time zero fields of the foliation determined by $L=1$ as well and 
$M,s_0$ are determined by $L$ via $e^{s_0 M}=L$. 

We can now combine (\ref{a.25}), (\ref{a.37}) and obtain 
\be \label{a.38}
H_L=O^L_h(0)=e^{s_0\{K_M,.\}}\cdot O^1_h(0)=e^{s_0\{K_M,.\}}\cdot H_1
\ee
We also may combine these to relate relational observables $O^L_f(t)$ to $O^{\hat{L}}_f(\hat{t})$ corresponding 
to a function $f$ of the true degrees of freedom by concatenation of canonical transformations
to obtain the relational reference frame transformation between different leaves, labelled by $z^0=t,\hat{z}^0=\hat{t}$ respectively of the 
different foliations defined by $L,\hat{L}$ respectively: Let $L=e^{s_0 M},\; \hat{L}=e^{\hat{s}_0 \hat{M}}$. We 
use the identity $[X_A,X_B]=X_{\{A,B\}}$ between Hamiltonian vector fields $X_A,X_B$ of functions $A,B$
and its implications
\be \label{a.39}
e^{s X_A}\cdot X_B \cdot e^{-s X_A}\cdot=X_{e^{s X_A}\cdot B}\cdot, \; 
e^{s X_A}\cdot e^{r X_B} \cdot e^{-s X_A}\cdot=\exp(r X_{e^{s X_A}\cdot B})\cdot
\ee
to find 
\ba \label{a.40}
O^L_f(t) &=& e^{t \{H_L,.\}}\cdot O^L_f(0)
=e^{t \{H_L,.\}}\cdot e^{s_0\{K_M,.\}}\cdot O^1_f(0)
=e^{s_0\{K_M,.\}}\cdot e^{t \{H_1,.\}}\cdot O^1_f(0)
\nonumber\\
&=& e^{s_0\{K_M,.\}}\cdot e^{t \{H_1,.\}}\cdot 
e^{-\hat{t} \{H_1,.\}}\cdot e^{-\hat{s}_0\{K_{\hat{M}},.\}}\cdot O^{\hat{L}}_f(\hat{t})
\ea
Note that all canonical transformations use only generators written in terms of relational 
observables of the time zero leaf of the $L=1$ foliation. If one wanted to write it in terms of the 
relational observables of the time zero fields of the $\hat{L}$ foliation we must go through 
the same derivation as above but use $L \hat{L}^{-1}=e^{s_0 M}$ instead. 
  
Let us work out (\ref{a.38}) for two examples, a rotation in the the $z^1=0$ plane and a 
boost in $z^1$ direction in $D=3$ corresponding to $M=R,\; M=B$ respectively. Using 
the indices $I,J\in \{0,1\},\;A,B\in \{2,3\}$  we have 
\be \label{a.41}
R^\mu_\nu=\delta^\mu_2\delta^3_\nu-\delta^\mu_3\delta^2_\nu,\;
B^\mu_\nu=\delta^\mu_0\delta^1_\nu+\delta^\mu_1\delta^0_\nu
\ee
and find 
\be \label{a.42}
\kappa_R=\int\; d^3 z\; \epsilon_{AB} \;z^A\;\delta^{BC} \pi\phi_{,C},\;
\kappa_B=\int\; d^3 z\ z^1\frac{1}{2}[\pi^2+\delta^{ab}\phi_{,a}\phi_{,b}+V(\phi)]
\ee
Then 
\be \label{a.43}
\{\kappa_R,h\}=0, \;\{\kappa_B,h\}=p:=\int\; d^3 z\; \pi\phi_{,1}(\vec{z}),\;
\{\kappa_B,p\}=h
\ee
which implies with $P:=p_{\phi\to O^1_\phi(0),\pi\to O^1_\pi(0)}$
\be \label{a.44}
H_R=H_1,\;H_B={\sf ch}(s_0) H_1+{\sf sh}(s_0)\;P
\ee
Thus as expected, if the frames are just rotated with respect to each other, then 
the two physical Hamiltonians coincide when pulled back by the the corresponding 
relational relative reference frame transformation because the Hamiltonian is rotation invariant, 
when they are boosted with 
respect to each other, then the two Hamiltonians do not coincide under the pullback,
rather one obtains a linear combination of Hamiltonian and momentum in boost direction.
This is due to the fact that the Hamiltonian is not a Lorentz scalar but rather 
the zero component of the energy momentum 4-vector. For completeness we record the 
generators of time and space translations, rotations and boosts in $D=3$ and 
any unit vector direction $\vec{n}$ given by $p_{\vec{n}}=n_a p^a$ and similar for 
$r_{\vec{n}},b_{\vec{n}}$ where 
\ba \label{a.45}
&& p^0=h,\; p^a=\int\;d^3 z\; \delta^{ab}\pi\phi_{,b},\;
b^a=m^0_a:=\int\; d^z\;z^a\frac{1}{2}(\pi^2+\delta^{ab}\phi_{,a}\phi_{,b}+V(\phi)](\vec{z}))
\nonumber\\
&& r^a=\epsilon^{abc} m_{bc}/2:=\int\; d^3 z\; \epsilon^{abc} \;\delta_{bd}\;z^d\;\pi\phi_{,c}(\vec{z}),\;
\ea
whose Poisson algebra with $m_{[\mu\nu)]}:=0$ we used in (\ref{a.43}) 
\be \label{a.46}
\{p_\mu,p_\nu\}=0,\;\{p_\mu,m_{\nu\rho}\}=2\eta_{\rho[\mu} p_{\nu]},\;
\{m_{\mu\nu},m_{\rho\sigma}\}=4 \eta_{[\mu[\rho} \;m_{\sigma]\nu]}
\ee
It closes for any $V$ and is isomorphic to the Poincar\'e algebra. \\
\\
The result (\ref{a.44}) can be obtained outside the context of parametrised field theory
by considering the Legendre transforms with respect to the $L=1,L\not=1$ foliations 
and the corresponding Hamiltonians $H_1,H_L$ and exploiting that both generate the same solutions of 
the Euler Lagrange equations. Then $H_L$ can be written as the restriction of a solution
to the time zero leaf of the $L\not=1$ foliation which can be written in terms of the 
time evolution of the initial data of that solution from the time zero leaf of the $L=1$ foliation.    
The above calculations
can be repeated in any $D$ and any $\vec{n}$ with appropriate changes. In the classical thery,
they can also 
be taken literally over to any other Poincar\'e covariant field theory (e.g. standard matter on 
Minkowski space) not only containing scalar
matter. They can also 
be taken over literally to the quantum theory for the case that $V$, or more generally the Lagrangian,
is at most quadratic in the fields (free field case)
because in that case the expressions (\ref{a.45}) can be defined as self-adjoint 
operators on the Fock space selected by $h$ after corresponding normal ordering. If $V$ is a higher order polynomial 
(\ref{a.45}) can only be defined as quadratic forms on the Fock space selected by the 
free part of $h$ and in that case a quantum treatment will require renormalisation.
In the free field case, the corresponding quantum relational reference transformation is simply 
the unitary representation of the (universal cover of the) Poincar\'e group on the field 
algebra.

\end{appendix}


\begin{thebibliography}{99}
    
\parskip -5pt    
                 
\bibitem{0} 
%A. Einstein. Albert Einstein and Michele Besso Correspondence 1903-1955. P. Speziali (ed.), Paris, Hermannn 1972
%(letter Jan. 1916)\\
%A. Einstein. 
%1961 Relativity and the problem of space. In:
%Relativity: the Special and General Theory. New York, Crown, 1961.\\
A. Einstein. In: Concepts of Space. M. Jammer (ed.). Harvard University Press, Cambridge, 1954.                
                 
\bibitem{1} P.G. Bergmann, A.B. Komar.
Poisson brackets between locally defined observables in general relativity.
Phys. Rev. Lett. 4 (1960) 432-433

\bibitem{3a} P. Mitra, R. Rajaraman. Gauge-invariant reformulation of an anomalous gauge theory.
Physics Letters B 225 (1989) 267–271.\\
R. Anishetty, A. S. Vytheeswaran. Gauge invariance in second-class constrained systems.
Journal of Physics A: Mathematical and General 26 (1993), no. 20 5613–5619.

\bibitem{3b} C. Rovelli. What is observable in classical and quantum gravity? Class. Quantum Grav. 8
(1991), 297-316.\\
C. Rovelli. Quantum reference systems. Class. Quantum Grav. 8 (1991), 317-332.

\bibitem{3c} B. Dittrich. Partial and complete observables for Hamiltonian constrained systems. Gen.
Rel. Grav. 39 (2007) 1891 [gr-qc/0411013].

\bibitem{3d} T. Thiemann. Reduced phase space quantization and Dirac observables. Class. Quant. Grav.
23 (2006), 1163-1180. [gr-qc/0411031].

\bibitem{3e} B. Dittrich. Partial and complete observables for canonical general relativity. Class. Quant.
Grav. 23 (2006),6155-6184. [gr-qc/0507106].\\
K. Giesel, T. Thiemann. Algebraic quantum gravity (AQG). IV. Reduced phase space quan-
tisation of loop quantum gravity. Class. Quant. Grav. 27 (2010) 175009. [arXiv:0711.0119
[gr-qc]].\\
M. Domagala, K. Giesel, W. Kaminski, J. Lewandowski. Gravity quantized: Loop Quantum
Gravity with a Scalar Field. Phys. Rev. D82 (2010) 104038. [arXiv:1009.2445 [gr-qc]]\\
V. Husain, T. Pawlowski. Time and a physical Hamiltonian for quantum gravity. Phys. Rev.
Lett. 108 (2012) 141301. [arXiv:1108.1145 [gr-qc]]\\
K. Giesel, T. Thiemann. Scalar Material Reference Systems and Loop Quantum Gravity
Class.Quant.Grav. 32 (2015) 135015. e-Print: 1206.3807 [gr-qc]
R. Ferrero, T. Thiemann. Asymptotically safe canonical quantum gravity: Gaussian dust matter.
e-Print: 2503.22474 [hep-th]

\bibitem{3f} Handbook of Quantum Gravity. C. Bambi, L. Modesto, I. Shapiro (eds.). 
Springer-Verlag, Berlin 2023.

\bibitem{3g} P. Hoehn, A.R.H. Smith, M.P.E. Lock. Trinity of relational quantum dynamics
Phys. Rev. {\bf D 104} (2021) 6, 066001. e-Print: 1912.00033 [quant-ph]

\bibitem{5} M. Henneaux and C. Teitelboim. Quantization of Gauge Systems. Princeton University
Press, Princeton, 1992.

\bibitem{4a} Aharonov, Y.; T. Kaufherr (1984). "Quantum frames of reference". Phys. Rev. D. 30 (2): 368–385. 

\bibitem{4b} F. Giacomini, E. Castro-Ruiz, C. Brukner. Quantum mechanics and the covariance of physical laws in quantum reference frames.
Nature Commun. 10 (2019) 1, 494. e-Print: 1712.07207 [quant-ph]

\bibitem{4c} S.D. Bartlett, T. Rudolph, R.W. Spekkens. Reference frames, superselection rules, and quantum information. 
Reviews of Modern Physics. 79 (2): 555–606. arXiv:quant-ph/0610030

\bibitem{4d} J.C. Fewster, D.W. Janssen, L.D. Loveridge, K. Rejzner, J. Waldron.
Quantum Reference Frames, Measurement Schemes and the Type of Local Algebras in Quantum Field Theory
Commun. Math. Phys. 406 (2025) 1, 19. e-Print: 2403.11973 [math-ph]

\bibitem{4e}  P.A. Hoehn, A. Russo, A.R.H. Smith. Matter relative to quantum hypersurfaces. 
Phys. Rev. {\bf D 109} (2024) 10, 105011. e-Print: 2308.12912 [quant-ph]

\bibitem{6} P. Dirac. Lectures on quantum mechanics. Dover books on physics, Dover 2001.

\bibitem{7} R. Wald. General Relativity. The University of Chicago Press, Chicago, 1984.
   
\bibitem{8} 
J.M. Pons, D.C. Salisbury, K.A. Sundermeyer.
Observables in classical canonical gravity: folklore demystified
J. Phys. Conf. Ser. 222 (2010) 012018. e-Print: 1001.2726 [gr-qc]\\
J.M. Pons, D.C. Salisbury, K.A. Sundermeyer.
Revisiting observables in generally covariant theories in the light of gauge fixing methods
Phys.Rev.D 80 (2009) 084015. e-Print: 0905.4564 [gr-qc]\\
J.M. Pons, D.C. Salisbury, K.A. Sundermeyer.
Gravitational observables, intrinsic coordinates, and canonical maps
Mod. Phys. Lett. A 24 (2009) 725-732. e-Print: 0902.0401 [gr-qc]\\
J.M. Pons, D.C. Salisbury.
The Issue of time in generally covariant theories and the Komar-Bergmann approach to observables in general relativity.
Phys. Rev. D 71 (2005) 124012. e-Print: gr-qc/0503013 [gr-qc]

\bibitem{8b} M.P. Woods, R. Silva, G. Puetz, S. Stupar, R. Renner. Quantum Clocks are More Accurate Than Classical Ones.
PRX Quantum 3 (2022) 1, 010319. e-Print: 1806.00491 [quant-ph]

\bibitem{8z} M. Bojowald, L. Martinez.
Large effects from quantum reference frames. J. Phys. A 58 (2025) 275302. e-Print:
2506.14721 [quant-ph]

\bibitem{8a} D. Giulini, D. Marolf.
On the generality of refined algebraic quantization. Class.Quant.Grav. 16 (1999) 2479-2488. e-Print:
gr-qc/9812024 [gr-qc]

\bibitem{8c} A.-C. de la Hamette, T.D. Galley, P.A. Hoehn, L. Loveridge, M.P. Mueller.
Perspective-neutral approach to quantum frame covariance for general symmetry groups
e-Print: 2110.13824 [quant-ph]

\bibitem{8d} A. Vanrietvelde, P.A. Hoehn, F. Giacomini, E. Castro-Ruiz.
A change of perspective: switching quantum reference frames via a perspective-neutral framework
Quantum 4 (2020) 225. e-Print: 1809.00556 [quant-ph]

\bibitem{9} C.G. Torre, M. Varadarajan. Quantum fields at any time. Phys. Rev. {\bf D 58} (1998) 064007.
e-Print: hep-th/9707221 [hep-th] \\
C.G. Torre, M. Varadarajan. Functional evolution of free quantum fields. Class. Quant. Grav. {\bf 16} (1999) 2651-2668.
e-Print: hep-th/9811222 [hep-th]

\bibitem{10} M. Montesinos, C. Rovelli, T. Thiemann. SL(2,R) model with two Hamiltonian constraints.
Phys.Rev.D 60 (1999) 044009. e-Print: gr-qc/9901073 [gr-qc]

\bibitem{11} C.J. Isham, A.C. Kakas.
A Group Theoretic Approach to the Canonical Quantization of Gravity. 1. Construction of the Canonical Group
Class. Quant. Grav. 1 (1984) 621\\
C.J. Isham, A.C. Kakas.
A Group Theoretical Approach to the Canonical Quantization of Gravity. 2. Unitary Representations of the Canonical Group.
Class. Quant. Grav. 1 (1984) 633

\bibitem{12} A. J. Hanson, T. Regge, C. Teitelboim. Constrained Hamiltonian systems.
Accademia Nazionale dei Lincei, 1976.

\bibitem{13} A. Ashtekar, R. Tate, C. Uggla. Minisuperspaces: Observables and quantization
Int. J. Mod. Phys.D 2 (1993) 15-50. e-Print: gr-qc/9302027 [gr-qc]

\bibitem{14} T. Thiemann. Reduced phase space induced decay decay conditions.

\bibitem{15} J. Neuser, T. Thiemann.
Quantum Field Theory of Black Hole Perturbations with Backreaction V. Beyond Second Order Perturbations
e-Print: 2602.08125 [gr-qc]

\bibitem{15a} C. Brukner, V. Kabel, W. Wieland. Quantum reference frames at the boundary of spacetime
Phys.Rev.D 108 (2023) 10, 106022. e-Print: 2302.11629 [gr-qc]\\
K. Giesel, V. Kabel, W. Wieland.
Linking edge modes and geometrical clocks in linearized gravity. Phys. Rev. D 112 (2025) 6, 064063. e-Print:
2410.17339 [gr-qc]

\bibitem{16} K. Kuchar. Geometry of Hyperspace. 1. J. Math. Phys. {\b 17} (1976) 777-791.\\
K. Kuchar. Kinematics of Tensor Fields in Hyperspace. 2. J. Math. Phys. {\bf 17} (1976) 792-800.
  
\bibitem{17} S.A. Fulling.  Aspects of Quantum Field Theory in Curved Spacetime. Cambridge University Press, Cambridge, 1989.

\bibitem{18} S.A. Hojman, K. Kuchar, C. Teitelboim. Geometrodynamics Regained. Annals Phys. 96
(1976) 88-135.




   
\end{thebibliography}
\end{document}